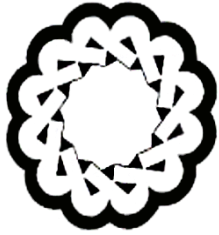

دانشکده علوم ریاضی و کامپیوتر

# تشخیص گفتار واج‌بنیان در زبان فارسی

گزارش پایان‌نامه‌ی کارشناسی ارشد

در رشته علوم کامپیوتر، گرایش محاسبات نرم و هوش مصنوعی

استاد راهنما:

دکتر محمدحسین قلی‌زاده

استاد مشاور:

دکتر سید ناصر رضوی

ارائه‌دهنده:

صابر ملک‌زاده

شهریور ۱۳۹۷

به پاس آنان که شرف را فدای نان و هدف را فدای جان نکردند، جان دادند تا از بند استثمار رهایی دهند و رهایی یابند،

در راه آزادی خون دادند، سوختند اما نساختند.

به پاس آنان که برای تعالی انسان کوشیدند.

به پاس پدربزرگم که تا آخرین نفس به راهم ایمان داشت و حمایتم کرد.

به پاس خانواده‌ام که مرا پرورش داده و همراهم بودند.

به پاس دکتر قلی‌زاده که در کسوت استادی مرا برادر بود.

به پاس دکتر رضوی که اهداف و مسیر زندگی را مدیون حضور ایشانم.

به پاس تمامی دوستان و آشنایانی که مرا به خاطر خودم، دوست داشته و همراهم بودند.

و در نهایت به پاس اساتید و راهنمایانی که عاشقانه و دلسوزانه از جوهره وجود خود در وجودم دمیدند.

# چکیده

به طور حتم یکی از مهم‌ترین موضوعات مطرح در علوم کامپیوتر، تشخیص گفتار هوشمند است. در این سامانه‌ها سعی می‌شود تا کامپیوترها نیز همانند انسان‌ها، بتوانند گفتارهایی که مخاطب‌شان قرار می‌گیرند را تشخیص داده و عکس‌العمل نشان دهند. در این پژوهش سعی شده تا با استفاده از ترکیب پردازش سیگنال و الگوریتم‌های کلاس‌بندی، روشی مناسب برای تشخیص واج‌های زبان فارسی توسط هوش مصنوعی کامپیوتری ارائه شود. برای این منظور برای پردازش سیگنال صوتی، از الگوریتم STFT و همچنین برای تشخیص و کلاس‌بندی سیگنال‌های پردازش‌شده از شبکه عصبی مصنوعی ژرف استفاده شده است. ابتدا نمونه‌های آموزشی به صورت دو واج‌های گفتار زبان فارسی تهیه شده و سپس عملیات پردازش سیگنال بر روی آن‌ها صورت گرفته است. سپس نتایج جهت آموزش داده‌ها به شبکه عصبی مصنوعی ژرف داده‌شده‌اند. در مرحله نهایی نیز عملیات آزمایش بر روی اصوات جدید صورت گرفته است.

**واژه‌های کلیدی:** تشخیص واج، تشخیص گفتار، یادگیری ژرف

# فهرست مطالب

# فهرست شکل‌ها

# فهرست جدول‌ها

# فصل ۱:

# مقدمه

## ۱-۱- مقدمه

در این فصل به بررسی چیستی مسئله و موضوع پژوهش پرداخته شده است. سپس تاریخچه‌ای از پژوهش‌های پیشین مرتبط در مورد سامانه‌های دانشگاه و تجاری تشخیص گفتار نیز ذکر شده است. درنهایت ساختار کلی گزارش مورد بررسی قرار گرفته است.

## ۱-۲- شرح مسئله

از مهم‌ترین کارها در تشخیص هر سیگنال صوتی، پردازش سیگنال و سپس تشخیص ساختار سیگنال‌های پردازش‌شده است. پردازش سیگنال همواره باهدف تشخیص ساختارهای جزئی موجود در سیگنال به جهت سادگی در تشخیص صورت می‌گیرد. زمانی که سیگنال پردازش می‌شود، ابتدا در حوزه‌های مختلف اعم از زمان و فرکانس و... بررسی می‌شود تا عملیاتی مانند حذف نویز یا استخراج ویژگی‌های مهم و موردنیاز در سیگنال و یا بسته‌بندی اطلاعات آن‌ها صورت پذیرد. پس از اینکه بر روی سیگنال عملیات پیش‌پردازش انجام شد، آماده است تا اطلاعات آن به جهت دسته‌بندی و تشخیص وارد الگوریتم‌های کلاس‌بندی و تشخیص‌دهنده شود.

در بخش تشخیص نیز الگوریتم‌های مختلفی مطرح هستند که در مرحله ابتدایی همگی سیگنال‌ها را در لایه ورودی دریافت کرده و متناسب با ویژگی‌های سیگنال تشخیص می‌دهند که هرکدام متعلق به کدام دسته‌ی از پیش تعیین‌شده‌ی خروجی هستند. پس‌ازاینکه الگوریتم آموخت که متناسب با ویژگی‌ها، سیگنال‌های ورودی را در دسته‌بندی‌های مناسب خروجی قرار دهد، می‌توان سیگنال‌هایی که دسته‌بندی آن‌ها مشخص نیست را به الگوریتم داده و کلاس خروجی متناسب با ویژگی‌های آن را از الگوریتم خواست.

از ابتدای زندگی هوشمندانه بشر بر روی زمین، وی همواره قصد داشته تا به‌وسیله شناسایی اصوات، در شناسایی محیط اطراف خود کوشا باشد. هرچند این تلاش تنها محدود به شناسایی اصوات محیط توسط انسان نبوده و عکس‌العمل محیط به صدای انسان نیز همواره مطرح بوده است. در این مسیر همواره تشخیص گفتار در راستای ارائه ورودی موردنظر انسان به کامپیوترها به‌وسیله صوت (همانند آنچه در میان انسان‌ها رایج است.) بسیار حائز اهمیت بوده است؛ چراکه سرعت انتقال اطلاعات به رایانه‌ها می‌توانست بسیار افزایش یابد.

## ۱-۳- تاریخچه

در زمینه تشخیص گفتار تلاش‌های فراوانی صورت گرفت و ابزارهای مکانیکی بسیاری نیز، قبل از شروع به کار ابزارآلات الکترونیکی در تاریخ بشر، تولید شد؛ اما شاید اولین محصول تولیدی در این زمینه که انقلابی شگرف به وجود آورد، گرامافون باشد که برای تولید اصوات از روی دیسک‌های اطلاعاتی وسط توماس ادیسون اختراع شد. [۱]

تشخیص خودکار گفتار به مدت پنجاه سال است که به‌صورت یک زمینه پژوهشی فعال، در حال پیشرفت است که همواره پلی برای ارتباط بهتر انسان با انسان و ماشین با انسان بوده است. البته در گذشته، تشخیص گفتار زمینه پژوهشی مهمی در ارتباط میان انسان و رایانه نبوده است و این موضوع به سه دلیل نبود کاربرد، محدودیت در استفاده و همچنین دقت پایین این سامانه‌ها بوده است.[۲] در سال‌های اخیر فناوری تشخیص گفتار شروع به تغییر روش زندگی ما نمود. بر طبق قانون Moore، کامپیوترها از لحاظ سرعت پردازش پیشرفت کردند که باعث شد متخصصان پردازش گفتار بتوانند مدل‌های پیچیده‌تری را برای تشخیص گفتار پیاده و اجرا کنند.[۳] نکته دوم مربوط به حافظه‌های بزرگ قابل‌استفاده برای ذخیره داده‌های حاصل از پردازش گفتار بود که با توجه به پیشرفت حافظه‌های ابری این امر نیز برای ما ممکن می‌شد. نکته دیگر گسترش دستگاه‌های همراه بود که با توجه به سختی امکان استفاده از صفحه‌کلید و اشاره‌گر در آن‌ها، نیاز به استفاده از گفتار برای ارائه دستوارت به آن‌ها با توجه به ماهیت گفتاری ارتباط میان انسان‌ها، بیش‌ازپیش دیده می‌شد.

اولین مسئله بر روی پردازش صدا جهت تشخیص گفتار، مسئله نمونه‌گیری از صداست. می‌دانیم صدا یک موج مکانیکی بوده و سیگنال صوتی یک سیگنال آنالوگ است. تشخیص سیگنال آنالوگ نیازمند سامانه‌های آنالوگ است که دارای پیچیدگی‌های بسیاری است. برای همین منظور استفاده از سیگنال گسسته با توجه به وجود الگوریتم‌های ریاضی مناسب برای بررسی ویژگی‌های مختلف سیگنال بسیار منطقی به نظر می‌رسد. برای تبدیل سیگنال آنالوگ که یک سیگنال مطلقاً پیوسته است به سیگنال گسسته از نمونه‌گیری استفاده می‌شود. از روش‌های مختلف برای نمونه‌گیری از سیگنال صوتی می‌توان به روش‌های periodic، cyclic rate، multirate، random و pulse width modulated اشاره کرد. از میان این روش‌ها، معمولاً از روش periodic استفاده می‌شود که درواقع نرخ نمونه‌گیری را در طول نمونه‌گیری ثابت در نظر می‌گیرد. [۴]

پس از نمونه گیری، سیگنال صدای حاصل در حوزه زمان حاوی اطلاعات محدودی است. برای اینکه ارتعاشات صدا و ویژگی‌های تغییرات صدا را بررسی کنیم با استفاده از تبدیلات ریاضی آن را به حوزه فرکانسی برده و ویژگی‌های فرکانسی را مورد بررسی قرار می‌دهیم. از مهم‌ترین بخش‌های موردنیاز جهت

بررسی، سازنده‌های فرکانس هستند. این سازنده‌ها که در نمودار زمان-فرکانس به‌صورت خطوط پررنگ دیده می‌شوند، درواقع شدت فرکانس‌های صدا در بازه‌های زمانی خاص را نشان می‌دهند. برای به دست آوردن نمودار زمان-فرکانس روش‌های متفاوتی وجود دارد که معمولاً از روش تبدیل فوریه زمان کوتاه استفاده می‌شود. تبدیل فوریه زمان کوتاه از اعمال تبدیل فوریه بر روی بازه‌های زمانی کوتاه از صدا به دست می‌آید. [۵]

پس از به دست آوردن شکل سیگنال در حوزه‌های مختلف به اعمال الگوریتم‌های مختلف پرداخته شده و تاریخچه بررسی سیگنال صوتی جهت تشخیص گفتار بیان می‌شود.

در سال ۱۹۳۲ در آزمایشگاه‌های تحقیقاتی بل، محققانی همچون Harvey Fletcher تحقیقات بر روی مفاهیم علمی گفتار را آغاز کردند. [۶] در سال ۱۹۵۲ محققان این آزمایشگاه موفق به ساخت سامانه‌ای شدند که بتواند اعداد بیان‌شده توسط یک گوینده‌ی خاص را تشخیص دهد. در دهه ۱۹۵۰ میلادی، اهم تلاش‌ها به تشخیص ده کلمه خلاصه می‌شد. [۷]

متأسفانه در سال ۱۹۶۹ شخصی به نام جان رابینسون پیرس نامه‌ای سرگشاده منتشر کرده و مدعی شد که تشخیص واج‌های زبان انگلیسی برای کامپیوترها غیرممکن است، بنابراین تشخیص گفتار برای کامپیوترها غیرممکن است. از همین رو تمام کمک‌های پژوهشی در آزمایشگاه‌های بل برای تشخیص گفتار قطع گردید. [۸] در اصل جان رابینسون راست می‌گفت. تشخیص واج به واج گفتار نه‌تنها برای کامپیوترها بلکه برای انسان نیز غیرممکن بود ولی جان رابینسون به این مسئله توجه نکرد که نیازی به تشخیص تک‌تک واج‌ها وجود ندارد و درواقع زمانی که انسان صحبت می‌کند برخی از واج‌ها را به خاطر افزایش سرعت سخن گفتن اصلاً بر زبان نمی‌آورد درحالی‌که طرف مقابل بااین‌وجود هم با درصد زیادی متوجه سخنان وی می‌شود که در ادامه توضیح داده خواهد شد.

در اواخر دهه ۱۹۶۰ میلادی دانشجوی هندی‌الاصل دانشگاه استنفورد، Raj Reddy موفق شد سامانه‌ای جهت تشخیص گفتار پیوسته ارائه دهد. تفاوت تشخیص گفتار پیوسته با گسسته در این بود که در تشخیص گفتار پیوسته کامپیوتر قادر بود جملات را تشخیص دهد درحالی‌که در تشخیص گفتار گسسته تنها کلمات جدا از هم به‌طوری‌که با مکث تلفظ شوند، تشخیص داده می‌شد. [۹] ۲۵ سال بعد، این دانشمند به دلیل تلاش‌هایش در علم هوش مصنوعی موفق به دریافت مهم‌ترین جایزه سالانه علوم کامپیوتر جهان، جایزه تورینگ شد.

در همین سال‌ها محققان اتحاد جماهیر شوروی موفق شدند با استفاده از الگوریتم کشش پویای زمان، روشی جهت تشخیص جملات به صورت پیوسته ارائه دهند. در این روش گفتار به بازه‌های ۱۰ میلی‌ثانیه‌ای

تقسیم می شد و هر بازه به صورت جداگانه پردازش می شد. هرچند بعدها دیگر از روش کشش پویای زمان استفاده نشد اما روش تقسیم گفتار به بازه‌های کوتاه همچنان مورداستفاده قرار می‌گرفت. [۱۰]

در سال ۱۹۷۱ موسسه تحقیقات دفاعی آمریکا، شروع به تغذیه‌ی مالی پروژه‌ای ۵ ساله باهدف تشخیص گفتار با ۱۰۰۰ کلمه انگلیسی کرد. در نتیجه محققانی از شرکت BBN، شرکت IBM، دانشگاه Carnegie Mellon و پژوهشگاه دانشگاه Stanford شروع به کار بر روی این پروژه کردند. [۱۱] [۱۲]

در اواخر دهه ۱۹۶۰ میلادی در موسسه آنالیز دفاعی، Leonard E. Baum بخش مدل ریاضی مدل زنجیره مارکوف را توسعه داد. سپس در دانشگاه Carnegie Mellon دانشجویان Raj Reddy به نام‌های James Baker و Janet Baker بر روی استفاده از مدل‌های پنهان مارکوف در تشخیص گفتار کارکردند. مدل پنهان مارکوف به پژوهشگران کمک کرد تا مدل‌های مختلف اصوات و زبان را مخلوط کرده و مدل احتمالی برای تشخیص گفتار ارائه دهند. [۱۳]

در اواسط دهه ۱۹۸۰ میلادی Fred Jelinek در شرکت IBM، توانست سامانه‌ای بسازد که بتواند بیش از ۲۰ هزار لغت را تشخیص داده و تایپ کند. این سیستم بیش از آنکه بر روی نحوه تشخیص گفتار توسط انسان تمرکز داشته باشد، بر روی مدل‌های تشخیص گفتار حاصل از مدل پنهان مارکوف تمرکز داشت. [۱۴]

در اوایل دهه ۱۹۹۰ میلادی یکی دیگر از دانشجویان Raj Reddy، به نام Xe Dong Hung، موفق شد sphinx ۲ را توسعه دهد که در سال ۱۹۹۲ موفق شد بهترین کارکرد را در بین تمامی سامانه‌های تشخیص گفتار موسسه تحقیقات دفاعی آمریکا داشته باشد. از مزیت‌های مهم این سیستم می‌توان به دایره لغات بالا، حذف نویز مناسب و همچنین حدس مناسب کلمات در جملات اشاره کرد. در سال ۱۹۹۶ میلادی نیز شرکت IBM اولین نرم‌افزار تشخیص گفتار پیوسته به نام MedSpeak را ارائه داد.

با شروع قرن ۲۱ میلادی سامانه‌های متعدد تجاری برای تشخیص گفتار منتشر شد و استفاده از شبکه عصبی در انواع مختلف آن برای تشخیص گفتار افزایش یافت.

در سال ۲۰۰۲ مایکروسافت تشخیص گفتار را به مجموعه Office اضافه کرد. [۱۵] در سال ۲۰۰۶ میلادی سازمان امنیت ملی آمریکا از تشخیص گفتار برای استخراج کلمات مهم گفتگوها استفاده نمود. [۱۶] در سال ۲۰۰۷ میلادی سیستم‌عامل Windows Vista ارائه شد که در آن برای اولین بار از تشخیص گفتار برای ارائه فرمان‌ها صوتی استفاده‌شده بود. در سال ۲۰۰۸ میلادی Google برای اولین بار سیستم تشخیص گفتار را به اولین گوشی هوشمند (Iphone) اضافه نمود. در سال ۲۰۱۱ اولین دستیار صوتی به نام Siri توسط Apple در گوشی‌های هوشمند این شرکت استفاده شد. در سال ۲۰۱۴ شرکت Microsoft از دستیار صوتی خود به نام Cortana رونمایی کرد و در سال ۲۰۱۵ میلادی Google اعلام نمود با استفاده از فناوری

شبکه‌های عصبی ژرف حافظه کوتاه-بلند مدت، توانسته بهبود فراوانی در این عرصه ایجاد کند.[۱۷]

## ۱-۴- ساختار گزارش

در فصل دوم این گزارش، مفاهیم مبنایی موردنیاز برای تفهیم مطالب موجود در توضیحات حوزه موضوع پایان‌نامه ارائه می‌شوند. فصل سوم شامل کارهای انجام شده در حوزه پایان‌نامه در گذشته است. در فصل چهارم روش پیشنهادی برای حل مسئله، ارائه و توضیح داده می‌شود و درنهایت در فصل پنجم نتایج پژوهش اعلام شده و با نتایج سایر کارها مقایسه می‌شوند.

# فصل ۲:

# تعاریف و مفاهیم مبنایی

## ۲-۱- مقدمه

هدف از این فصل ارائه برخی مفاهیم مبنایی از حوزه‌های پردازش سیگنال و همچنین یادگیری ماشینی، جهت آشنایی خواننده و آمادگی وی جهت بررسی مفاهیم پردازش صوت و گفتار است.

## ۲-۲- پردازش سیگنال

سیگنال به موجی گفته می‌شود که حاوی اطلاعاتی در شکل امواج آنان باشد. سیگنال‌های دیجیتالی به سیگنال‌هایی گفته می‌شوند که به‌صورت دیجیتالی ذخیره یا به نمایش گذاشته می‌شوند. در این میان برای نمایش اطلاعاتی مانند صوت یا تصویر می‌توانیم آن‌ها را به‌صورت امواج صوتی یا تصویری به نمایش درآوریم. پردازش سیگنال دیجیتال یعنی بررسی دیجیتالی امواج باهدف تشخیص خصوصیات وجودی آن‌ها. برای پردازش سیگنال الگوریتم‌های متفاوتی موجود است که ویژگی‌های مهم و خاص سیگنال‌ها را برجسته کرده و آن‌ها را آشکار می‌سازد و بدین‌وسیله خواهیم توانست ویژگی‌های متمایزکننده یک سیگنال از سایر سیگنال‌ها را بررسی کنیم.

همان‌گونه که بیان شد، سیگنال حاصل از تغییرات موج همواره دارای فرکانس یا ارتعاشات در واحد زمان است. به جهت پی بردن به ماهیت فرکانسی سیگنال باید تغییرات آن در حوزه فرکانسی نیز بررسی شود. به همین جهت تبدیلات ریاضی مختلفی مورد بحث قرار گرفته است که در ادامه توضیح داده خواهد شد.[۱۸]

## ۲-۳- تبدیلات ریاضی مطرح در پردازش سیگنال

### ۲-۳-۱- تبدیل موجک

تبدیل موجک نیز یک تبدیل ریاضی است که سیگنال را از حوزه زمانی به حوزه زمان-فرکانس می‌برد.

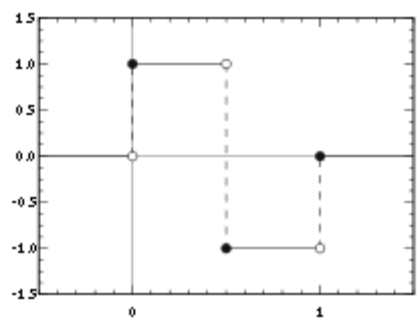

شکل ۲-۱- موجک هار

ساده‌ترین نوع آن موجک هار است که در شکل (۲-۱)، شکل این موجک دیده می‌شد.[۱۹]

## ۲-۳-۲- تبدیل فوریه

تبدیل فوریه به ا سم ریا ضیدان فران سوی ژوزف فوریه، ارائه‌دهنده این تبدیل نام‌گذاری شده ا ست. این تبدیل یک تابع انتگرالی است که هر تابع $x(t)$ را به تابع $X(f)$ منعکس می‌کند و رابطه آن به شرح زیر است:

$$X(f) = \int_{-\infty}^{+\infty} x(t)\, e^{-j2\pi f_t}\, dt$$

در حوزه پردازش سیگنال می‌توان با اعمال تبدیل فوریه، هر سیگنال را به حوزه فرکانسی برد. مقادیر تبدیل فوریه، مقادیر سیگنال در حوزه فرکانسی هستند.[۲۰]

### ۲-۳-۲-۱- تبدیل فوریه زمان کوتاه

تبدیل فوریه زمان کوتاه نوعی از تبدیل فوریه است که در بازه‌های زمانی متفاوت از سیگنال تبدیل فوریه می‌گیرد. در این تبدیل ابتدا سیگنال به بازه‌های زمانی کوتاه تقسیم‌بندی شده و سپس از سیگنال‌های تقسیم‌بندی شده تبدیل فوریه گرفته می‌شود. تبدیل فوریه زمان کوتاه برای استخراج فرکانس‌های سیگنال در بازه‌های زمانی متفاوت ا ستفاده می شود. این تبدیل یک ماتریس با مقادیر زمانی و فرکان سی می‌دهد که می‌توان از آن شدت هر فرکانس در بازه‌های زمانی مختلف را به دست آورد. [۲۱]

$$\mathbf{STFT}\{x(t)\}(\tau,\omega) \equiv X(\tau,\,\omega) = \int_{-\infty}^{+\infty} x(t) w(t-\tau) e^{-j\omega t}\, dt$$

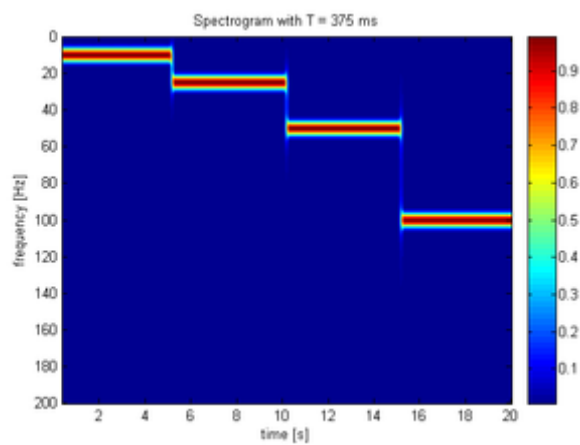

شکل ۲-۲- نمونه نمودار **STFT**

در فرمول بالا، W تابع پنجره، x سیگنال مورد بررسی، ω بازه زمانی و τ بازه زمانی است.

## ۲-۳-۳- Mel-frequency Cepstrum

تبدیلی است حاصل از ترکیب دو تبدیل فوریه و کسینوس که حاصل آن، سیگنال را در حوزه زمان-فرکانس نمایش می‌دهد. مراحل انجام عملیات مربوط به آن به شرح زیر است:

۱- اعمال تبدیل فوریه

۲- تبدیل توان‌های موجود در نمودار Spectrum به مقیاس Mel، به‌وسیله پنجره‌های متداخل مثلثی

۳- گرفتن لگاریتم از هرکدام از توان‌های موجود در مقیاس Mel فرکانسی

۴- اعمال تبدیل کسینوس بر نتیجه‌ی مورد ۳

با استفاده از این تبدیل، ماتریس و درنتیجه نموداری سه‌بعدی به دست می‌آید که می‌توان از آن برای استخراج ویژگی‌های صوت استفاده کرد. [۲۲]

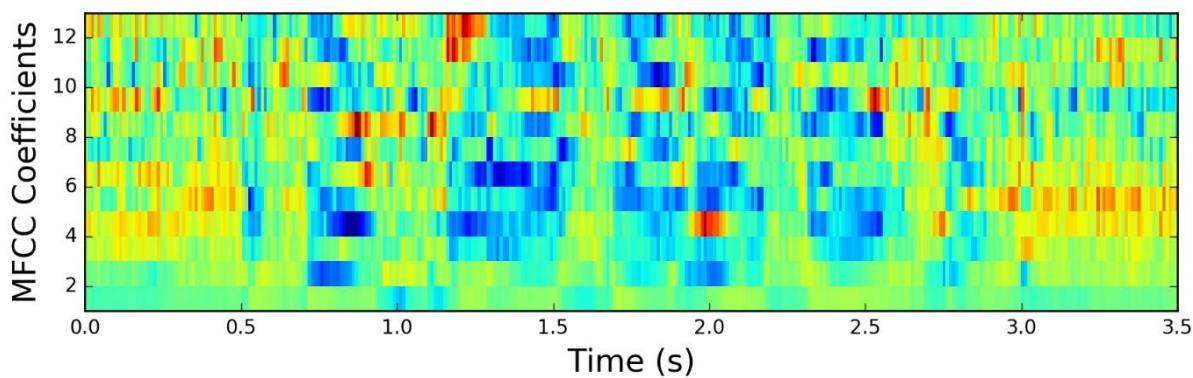

شکل ۲-۳- نمونه نمودار MFCC

## ۲-۴- تشخیص اصوات

### ۲-۴-۱- تفاوت در اصوات

در اصل تفاوت در اصوات به تفاوت در تغییرات شدت صدا و همچنین تفاوت فرکانس‌های ظاهرشده در سیگنال آن است. شدت صدا درواقع میزان بلندی و کوتاهی صدا و فرکانس آن میزان ارتعاشات امواج صوتی در واحد زمان است. هرکدام از این ویژگی‌ها می‌تواند به ما در تشخیص ویژگی‌های مختلف صوت یاری رساند. [۲۳]

### ۲-۴-۲- حوزه زمانی

بررسی سیگنال در حوزه زمانی به معنای بررسی شدت یا همان بلندی و کوتاهی سیگنال صوتی است. در این حوزه اطلاعاتی مستقیم از فرکانس در دست نیست، اما می‌توان با استفاده از تبدیلات ریاضی سیگنال را از حوزه زمان به حوزه فرکانسی انتقال داد. در اصل در تمام حوزه‌ها اطلاعاتی از صوت وجود دارد که می‌توان با تبدیل حوزه‌ها به یکدیگر آن اطلاعات را استخراج کرد. عملیات تبدیل حوزه‌ها به هم نه چیزی به سیگنال اضافه و نه چیزی از آن کم می‌کند. این کار صرفاً تبدیل‌کننده حوزه‌ها به یکدیگر و برجسته کردن برخی ویژگی‌ها در حوزه مورد بررسی است. [۲۳]

### ۲-۴-۳- حوزه فرکانسی

در این حوزه اطلاعات فرکانسی سیگنال بررسی می‌شود. در حوزه فرکانسی نمودارها به‌صورت فرکانس و تعداد دفعات ظهور فرکانس در سیگنال بررسی می‌شوند. [۲۳]

### ۲-۴-۴- سختی تشخیص اصوات

تشخیص حتمی اصوات، از آن روی که باید بدون خطا باشد، امری قطعاً غیرممکن است؛ اما در حوزه تشخیص اصوات هدف، یافتن جزئیات اصوات برای استخراج ویژگی‌های آنان و درنتیجه متمایز کردن اصوات از یکدیگر است. جزئیات صوت شدت صوت و فرکانس آن است اما با کمی تغییر در شدت و فرکانس، ممکن است قدرت تصمیم‌گیری برای متمایز کردن اصوات بسیار کم شود و این تغییرات تشخیص را سخت‌تر می‌کند. [۲۴]

## ۲-۵- تشخیص گفتار

### ۲-۵-۱- سختی تشخیص گفتار

در زمان ساخت اولیه یک زبان، همواره معیار مشخص کردن واج‌ها، کار راحتی است؛ اما همین راحتی تلفظ در زبان باعث سختی تشخیص گفتار می‌شود چراکه واج‌ها نزدیک به هم بوده و در اکثر زبان‌ها تمایز

زیادی با یکدیگر ندارند. برای مثال حروف «ب» و «پ» از نظر تلفظ بسیار نزدیک به هم هستند. استفاده ترکیب این دو واج بهجای هم که از نظر تلفظ بسیار به هم نزدیک‌اند برای سامانه‌های تشخیص گفتار بسیار محتمل است چراکه هم از نظر نوع تلفظ و هم از نظر فرکانس بسیار نزدیک به هم‌اند و معدود معیارهای تشخیص گفتار را به اشتباه می‌اندازد.

همچنین شباهت در فرکانس و بی‌تأثیر بودن معیار شدت صدا در تشخیص واج‌های مشابه گفتار، تشخیص را سخت‌تر می‌کند.

### ۲-۵-۲- مشکلات تشخیص

سختی تشخیص گفتار در شباهت فرکانس واج‌ها و همچنین سرعت تلفظ است. زمانی که واج‌ها با سرعت بالا پشت سرهم تلفظ می‌شوند، بعضاً برخی از آن به‌کلی تلفظ نمی‌شوند یا زمان و شدت تلفظ‌ها برای واج‌های مختلف با هم متفاوت است و این‌ها مشکلاتی هستند که در تشخیص گفتار باید در نظر داشت.

برای تشخیص واج‌های افتاده در گفتار می‌توان از میزان شباهت بخش تلفظ شده کلمه با دیکشنری زبان برای تشخیص کامل کلمه تلفظ شده است. [۲۵] این امر مشابه آن چه می باشد که در مغز انسان رخ می دهد. همچنین به عنوان یکی از بهترین روش‌ها برای تشخیص تفاوت‌های ریز فرکانسی برای پردازش هر چه بهتر سیگنال صوتی می‌توان استفاده کرد.

تفاوت در زبان‌ها می‌تواند مشکل‌ساز باشد، چراکه باعث تغییر واج‌های اساسی زبان و همچنین نحوه تلفظ واج‌ها می‌شود و این تمایز میان اصوات، درنتیجه تشخیص اصوات گفتار را سخت‌تر می‌نماید چراکه هر سیستم تشخیص گفتار برای زبان خاصی تمرین داده می‌شود و همچنین با استفاده از دیکشنری زبان خاصی کار می‌کند. [۲۵]

## ۲-۶- شبکه‌های عصبی مصنوعی

### ۲-۶-۱- ایده اصلی و موارد استفاده

ایده اصلی شبکه‌های عصبی مصنوعی از یادگیری انسان به‌وسیله‌ی ارتباط بین سلول‌های عصبی مغزی ا ست. در مغز ان سان هر نورون به صورت م ستقل عملیات پردازش و ذخیره سازی اجرایی داده‌ها را بر عهده

دارد. در شبکه‌های عصبی مصنوعی نیز در لایه‌های شبکه، نورون‌هایی قرار دارند که داده‌ها را دریافت کرده و به پردازش آن‌ها جهت شناسایی ساختارشان می‌پردازند.

موارد مصرف شبکه‌های عصبی، پردازش اطلاعات جهت تشخیص الگوهایی در آن‌ها جهت تشخیص نوع آن داده‌هاست. برای مثال می‌توان از شبکه‌های عصبی جهت تشخیص اینکه یک سیگنال صدا دارای چه ویژگی‌هایی است، چگونه می‌توان این ویژگی‌ها را استخراج کرد و با توجه به این ویژگی‌ها در چه نوعی از صداها دسته‌بندی می‌شود می‌توان استفاده کرد. [۲۶]

### ۲-۶-۲- یادگیری ژرف

یادگیری ژرف زیرشاخه‌ای از یادگیری ماشینی است که از لایه‌های متعدد تبدیلات خطی به‌منظور پردازش سیگنال‌های حسی مانند صدا و تصویر استفاده می‌کند. ماشین در این روش هر مفهوم پیچیده را به مفاهیم ساده‌تری تقسیم می‌کند و با ادامه این روند به مفاهیم پایه‌ای می‌رسد که قادر به تصمیم‌گیری برای آن‌ها است و بدین ترتیب نیازی به نظارت کامل انسان برای مشخص کردن اطلاعات لازم ماشین در هر لحظه نیست. موضوعی که در یادگیری ژرف اهمیت زیادی دارد، نحوه ارائه اطلاعات است. ارائه دادن اطلاعات به ماشین باید به نحوی باشد که ماشین در کمترین زمان اطلاعات کلیدی را که می‌تواند با استناد به آن‌ها تصمیم بگیرد را دریافت کند. هنگام طراحی الگوریتم‌های یادگیری ژرف می‌بایست به عوامل دگرگونی[1] که اطلاعات مشاهده‌شده را توضیح می‌دهند توجه کنیم. این عوامل معمولاً عوامل قابل مشاهده‌ای نیستند بلکه عواملی هستند که بر روی دسته قابل مشاهده تأثیرگذار بوده یا زاده ساختارهای ذهنی انسان برای ساده‌تر کردن مسائل هستند. برای مثال در هنگام پردازش گفتار، عوامل دگرگونی می‌توانند لهجه گوینده، سن یا جنسیت او باشند. در هنگام پردازش تصویر یک ماشین، میزان درخشش خورشید یک عامل دگرگونی است. یکی از مشکلات هوش مصنوعی تأثیر زیاد عوامل دگرگونی بر روی اطلاعات دریافتی است. برای مثال بسیاری از پیکسل‌های دریافتی از یک ماشین قرمز در شب ممکن است سیاه دیده بشوند. برای حل این مشکلات بعضاً به درک بالای اطلاعات (در حدود انسان) نیازمندیم و درواقع گاهی یافتن نحوه مناسب نمایش اطلاعات به اندازه خود مسئله سخت و زمان‌بر است. از مهم‌ترین انواع شبکه‌های مصنوعی ژرف می‌توان به شبکه‌های کانولوشن[2] و انواع شبکه‌های رکارنت[3] اشاره کرد.

---

[1] factors of variation
[2] Convolution
[3] Recurrent

شبکه‌های کانولوشن متشکل از تعدادی لایه کانولوشن، تعدادی لایه پولینگ[1] و تعدادی لایه تمام متصل[2] هستند. از این شبکه‌ها معمولا در دسته‌بندی تصاویر و یا به طور کلی استخراج ویژگی‌ها از تصاویر استفاده می‌شود؛ اما این تنها کاربرد این شبکه‌ها نیست. شبکه‌های کانولوشن نه‌تنها تصاویر بلکه قادر هستند هر نوع داده‌هایی را در ابعاد مختلف مورد بررسی قرار داده و ویژگی‌های موجود در آن‌ها را با سرعت بالا یافته و پردازش کنند. این شبکه‌ها برای یافتن ویژگی‌ها بدون در نظر گرفتن محل وقوع ویژگی بسیار مناسب هستند. همچنین از مهم‌ترین ویژگی‌های این شبکه‌ها می‌توان به توانایی آن‌ها در استخراج ویژگی‌های چندبعدی و به عبارتی غیرخطی اشاره کرد.

در کنار شبکه‌های کانولوشن شبکه‌های ریکارنت قرار دارند. این شبکه‌ها قادرند ویژگی‌ها را با در نظر گرفتن و بررسی محل وقوع آن‌ها و همچنین احتساب سایر ویژگی‌های رخ داده در نزدیکی محل وقوع آن‌ها، عملیات استخراج ویژگی را انجام دهند. این ویژگی‌ها زمانی مهم می‌شوند که احتمال وقوع یک ویژگی در داده ورودی کاملاً وابسته به‌احتمال حضور یک ویژگی دیگر در آن داده است. شبکه‌های ریکارنت ساختاری یک‌بعدی و به عبارتی خطی دارند بدین معنا که داده‌های ورودی به این شبکه‌ها باید از نوع داده‌های یک‌بعدی و به‌صورت دنباله باشند.

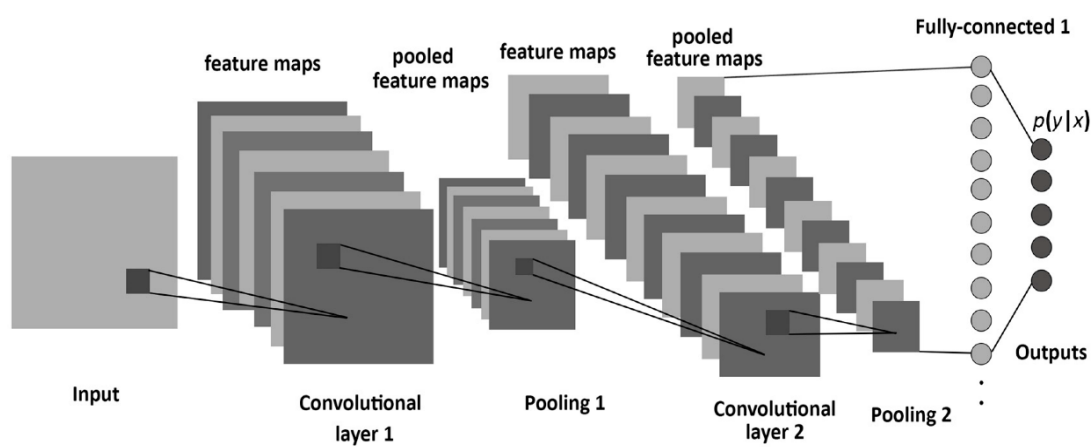

شکل ۲-۴- معماری شبکه عصبی کانولوشن[۲۷]

در عمل معمولا شبکه‌های ریکارنت برای استخراج ویژگی‌هایی به کار می‌روند که دارای ویژگی‌های مرتبط در حوزه سری زمانی باشند. برای مثال اگر در وسط جمله یک جای خالی برای حضور یک کلمه ایجاد کنیم، پر کردن آن جای خالی با توجه به اطلاعات زبانی کاملاً وابسته به سایر کلمات موجود در جمله است. از همین رو این شبکه‌ها در استخراج ویژگی از سری‌های زمانی یا دنباله‌های خطی بسیار مناسب عمل

---
[1] Pooling
[2] Fully Connected

می‌کنند. در کنار این شبکه‌ها، شبکه‌های کانولوشن یک دید کلی از وجود یا عدم وجود ویژگی‌ها در داده‌های ورودی می‌دهند که این عملیات بدون توجه به ابعاد داده صورت می‌گیرد. در کل می‌توان گفت با توجه به سرعت و جامعیت شبکه‌های کانولوشن، در صورتی که ترتیب حضور ویژگی‌ها در داده‌ها مهم نباشد، استفاده از شبکه‌های کانولوشن به‌احتمال بسیار، منجر به نتایج بهتری می‌گردد.

# فصل ۳:

# مروری بر کارهای گذشته

## ۳-۱- مقدمه

هدف از این فصل بررسی کارهای گذشته انجام پذیرفته در حوزه موضوع پایان‌نامه و نقاط قوت و ضعف آن‌هاست. در این فصل به بررسی سامانه‌های پردازش صوت و به طور خاص پردازش گفتار خواهیم پرداخت و پیشرفت‌ها و کاستی‌های موجود در این سامانه‌ها را بررسی خواهیم کرد.

## ۳-۲- پردازش سیگنال

### ۳-۲-۱- پردازش سیگنال دیجیتال به شکل امروزی

تقریباً از آغاز گسترش رایانه‌ها به شکل امروزی (دیجیتالی) پردازش سیگنال نیز وارد دنیای رایانه‌ها شد. در این زمینه سیگنال‌های مکانیکی و الکترومغناطیسی سی اعم از صوت، تصویر، سیگنال‌های مغزی و... مورد بررسی قرارگرفته‌اند.

جهت تشخیص ویژگی‌های سیگنال‌ها از تبدیلات ریاضی متنوعی مانند فوریه، موجک، مانند Curvelet و Ridglet استفاده می‌شود. این تبدیلات کمک می‌کنند تا ویژگی‌های مختلف سیگنال‌ها اعم از زمانی و فرکانسی و... مورد بررسی قرار گیرند.[۲۸]

### ۳-۲-۲- پردازش سیگنال در پردازش صوت

صوت به‌عنوان نوعی از سیگنال، به جهت برجسته‌سازی اطلاعات مهم و استخراج ویژگی می‌تواند توسط رایانه‌ها مورد پردازش قرار گیرد. با توجه تک‌بعدی بودن صوت در واحد زمان، پردازش صوت نیازمند پردازش‌های سبک تری خواهد بود. اما با توجه به ریز بودن تفاوت‌های اصوات و موجود بودن تنها یک پارامتر (شدت) برای تمایز اصوات، تشخیص اصوات را بسیار سخت‌تر کرده است. از این جهت در کارهای گذشته همواره سعی بر این بوده است تا با استفاده از الگوریتم‌های مختلف پردازش سیگنال، حداکثر ویژگی‌های ممکن در تمامی حوزه‌ها برای راحتی تشخیص از اصوات استخراج شوند.[۱۸]

## ۳-۳- شبکه عصبی مصنوعی

### ۳-۳-۱- شبکه عصبی پرسپترون[1] چندلایه

شبکه عصبی مصنوعی پرسپترون ساده‌ترین نوع شبکه‌های عصبی مصنوعی است که در سال ۱۹۵۷ به‌صورت رسمی ارائه شد. در این شبکه متغیر در وزن‌ها ضرب می‌شد و با یک بایاس (منحرف‌کننده به سمت تعادل سیستم) ترکیب‌شده و مقدار آن درصورتی‌که بزرگ‌تر از صفر بود در مکانی که متغیر ورودی قرار داشت قرار می‌گرفت. در مرحله بعدی وزن‌ها به صورتی که خطای گرادیان را کاهش دهند حرکت می‌کردند. تعداد تکرارها بستگی به خطای گرادیان موردنظر داشت. پس از مدتی شبکه‌های پرسپترون چندلایه نیز معرفی شدند که در هر تکرار چند بار عملیات تحلیل متغیرها انجام می‌شد و این باعث می‌شد پیش رفتن به‌سوی وزن‌هایی که خطای گرادیان را کاهش می‌دهند با دقت بیشتری انجام شود. [۲۹]

### ۳-۳-۲- یادگیری ژرف

در سال ۱۹۷۱ با بررسی شبکه‌های عصبی چندلایه، یادگیری ژرف نیز متولد شد. در این سال دانشمندی به نام الکسی مقاله‌ای منتشر ساخت که در آن، به بررسی شبکه‌ای ۸ لایه پرداخته و از آن استفاده کرده بود. وی این شبکه را شبکه عصبی ژرف نامید چراکه با استفاده از لایه‌های بسیار آن نسبت به شبکه‌های عصبی گذشته، می‌توانست جزئیات بیشتری را از تصویر ورودی بررسی کند. در اصل افزایش تعداد هرچند سرعت را پایین می‌آورد و به‌تبع آن دقت را نیز کاهش می‌داد، اما جزئیات را به‌صورت دقیق‌تری از سیگنال ورودی بررسی می‌کرد.

در سال ۲۰۰۱ به‌صورت رسمی اولین محصول رسمی تشخیص گوینده با استفاده از شبکه عصبی منتشر شد که می‌توانست صدای گوینده‌های مختلف را از یکدیگر تشخیص دهد. همچنین این شبکه می‌توانست تعدادی از لغات تلفظ شده را نیز متوجه شود و این شروعی به تجاری‌سازی سامانه‌های تشخیص گفتار بود. در سال‌های اخیر، با گسترش مفاهیم یادگیری ژرف در میان پژوهشگران این عرصه و همچنین ارائه معماری‌های نوین بر پایه‌ی یادگیری ژرف، مانند شبکه‌های عصبی کانولوشن و ریکارنت، تشخیص گفتار نسبت به قبل بسیار آسان‌تر شده است، اما هنوز چالش‌های مختلفی در این زمینه مطرح است که رسیدن به تشخیصی همانند تشخیص انسان را سخت کرده است. امید است تا باوجود یادگیری ماشین‌ها و بهبود مداوم عملکرد آن‌ها، بهبود فراوانی در تشخیص گفتار در سال‌های آتی را نیز شاهد باشیم. [۲۶]

از قرن نوزدهم به طور هم‌زمان اما جداگانه از سویی نروفیزیولوژیست‌ها سعی کردند سیستم یادگیری و

---

[1] Perceptron

تجزیه‌وتحلیل مغز را کشف کنند و از سوی دیگر ریاضیدانان تلاش کردند مدل ریاضی بسازند که قابلیت فراگیری و تجزیه‌وتحلیل عمومی مسائل را دارا باشد. اولین کوشش‌ها در شبیه‌سازی با استفاده از یک مدل منطقی در اوایل دهه ۱۹۴۰ توسط وارن مک‌کالک[1] و والتر پیتز[2] انجام شد که امروزه بلوک اصلی سازنده اکثر شبکه‌های عصبی مصنوعی است. عملکرد این مدل مبتنی بر جمع ورودی‌ها و ایجاد خروجی با استفاده از شبکه‌ای از نورون‌ها است. اگر حاصل جمع ورودی‌ها از مقدار آستانه بیشتر باشد، اصطلاحاً نورون[3] برانگیخته می‌شود. نتیجه این مدل اجرای ترکیبی از توابع منطقی بود. در سال ۱۹۴۹ دونالد هب[4] قانون یادگیری را برای شبکه‌های عصبی طراحی کرد. در سال ۱۹۵۸ شبکه پرسپترون توسط روزنبلات[5] معرفی گردید. این شبکه نظیر واحدهای مدل شده قبلی بود. پرسپترون دارای سه لایه است که شامل لایه ورودی، لایه خروجی و لایه میانی می‌شود. این سیستم می‌تواند یاد بگیرد که با روشی تکرارشونده وزن‌ها را به‌گونه‌ای تنظیم کند که شبکه توان بازتولید جفت‌های ورودی و خروجی را داشته باشد. روش دیگر، مدل خطی تطبیقی نورون است که در سال ۱۹۶۰ توسط برنارد ویدرو[6] و مارسیان هاف[7] در دانشگاه استنفورد به وجود آمد که اولین شبکه‌های عصبی به کار گرفته شده در مسائل واقعی بودند. آدالاین[8] یک دستگاه الکترونیکی بود که از اجزای ساده‌ای تشکیل شده بود و روشی که در آن برای آموزش استفاده می‌شد با پرسپترون فرق داشت. در سال ۱۹۶۹ مینسکی[9] و پاپرت[10] کتابی نوشتند که محدودیت‌های سامانه‌های یک‌لایه و چندلایه پرسپترون را تشریح کردند. نتیجه این کتاب پیش‌داوری و قطع سرمایه‌گذاری برای تحقیقات در زمینه شبیه‌سازی شبکه‌های عصبی بود. آن‌ها با طرح اینکه طرح پرسپترون قادر به حل هیچ مسئله جالبی نیست، تحقیقات در این زمینه را برای مدت چندین سال متوقف کردند. باوجوداینکه اشتیاق عمومی و سرمایه‌گذاری‌های موجود به حداقل خود رسیده بود، برخی محققان تحقیقات خود را برای ساخت ماشین‌هایی که توانایی حل مسائلی از قبیل تشخیص الگو را داشته باشند، ادامه دادند. ازجمله گراسبگ[11]

---

که شبکه‌ای تحت عنوان آوالانچ[1] را برای تشخیص صحبت پیوسته و کنترل دست ربات مطرح کرد. همچنین او با همکاری کارپنتر[2] شبکه‌های نظریه تشدید انطباقی را بنا نهادند که با مدل‌های طبیعی تفاوت داشت. اندرسون[3] و کوهونن[4] نیز از اشخاصی بودند که فن‌هایی برای یادگیری ایجاد کردند. ورباس[5] در سال ۱۹۷۴ شیوه آموزش پس انتشار خطا را ایجاد کرد که یک شبکه پرسپترون چندلایه البته با قوانین نیرومندتر آموزشی بود. پیشرفت‌هایی که در سال ۱۹۷۰ تا ۱۹۸۰ به دست آمد برای جلب‌توجه به شبکه‌های عصبی بسیار مهم بود. برخی فاکتورها نیز در تشدید این مسئله دخالت داشتند. امروز نیز تحولات زیادی در فن‌آوری شبکه‌های عصبی مصنوعی ایجادشده است. در دهه ۹۰ میلادی دانشمندان هوش‌مصنوعی سعی کردند تا با استفاده از مفهوم شبکه‌های عصبی، انواع شبکه‌های عصبی را توسعه دهند. از انواع این شبکه‌ها می‌توان به شبکه‌های عصبی مصنوعی ژرف اشاره کرد که در زمان خود کمک شایانی به بهبود شرایط و پیشبرد الگوریتم‌ها در زمینه‌های مختلف داشته‌اند.

از سال ۲۰۱۲ با Alexnet شبکه‌های عصبی شکل تازه‌ای به خود گرفتند. تعداد لایه‌ها به بیشتر از سه لایه افزایش پیدا کرده و مفهوم شبکه‌های عصبی ژرف ایجاد شدند. این شبکه‌ها معماری‌های مختلفی داشتند که حاصل از کاربرد و یا بهینه بودن نتایج در انواع مختلف آنان بود.

همچنین در سال‌های اخیر این شبکه‌ها نه‌تنها معماری‌های مختلفی داشته‌اند که از آن جمله می‌توان به Resnet، Densnet و Nasnet اشاره کرد، بلکه ساختارهای مختلفی برای کاربردهای مختلف نیز ارائه شدند. برای مثال از این شبکه‌ها برای تشخیص تفاوت‌های ساختاری بین تصاویر مانند تشخیص چهره، تفاوت‌های معنایی در متون مانند استخراج معنا، ترجمه ماشینی، قطعه‌بندی تصاویر، تشخیص اشیا و مهم‌ترین آن‌ها یعنی کلاس‌بندی تصاویر، متون و... استفاده کرد.

### ۳-۳-۳- علل بهبود تشخیص

تشخیص اصوات به دلیل وجود جزئیات بسیار و اطلاعات کم همواره بسیار مشکل بوده است. ازاین‌رو همواره سعی شده بهترین روش‌ها برای این منظور در نظر گرفته شود. در بسیاری از آزمایش‌ها شبکه عصبی مصنوعی با توجه به ساختار شبیه به ساختار مغز آن و همچنین الگوریتم‌های بهینه و مناسب، همواره بهترین

---

[1] Avalanch
[2] Gail Carpenter
[3] James Anderson
[4] Teuvo Kohonen
[5] Verbose

نتیجه را ارائه داده است.

## ۳-۴- پردازش گفتار

### ۳-۴-۱- ویژگی‌های گفتار

گویش انسان همانند سایر اصوات موجی مکانیکی است که حاصل ارتعاشات حنجره، میزان فشار به حنجره جهت خروج صدا و همچنین تغییرات شکل دهان و زبان و دندان‌ها جهت تغییر شکل ادای واج‌هاست. سختی تشخیص گفتار زمانی روشن می‌شود که کمی تغییر زاویه در شکل زبان و دندان‌ها باعث تغییر اصوات می‌شوند. این تغییرات به‌قدری کوچک‌اند که یافتن این تغییرات در شکل سیگنال‌ها، هم در حوزه زمانی، هم در حوزه فرکانسی و هم در سایر حوزه‌ها بسیار مشکل است. [۳۰]

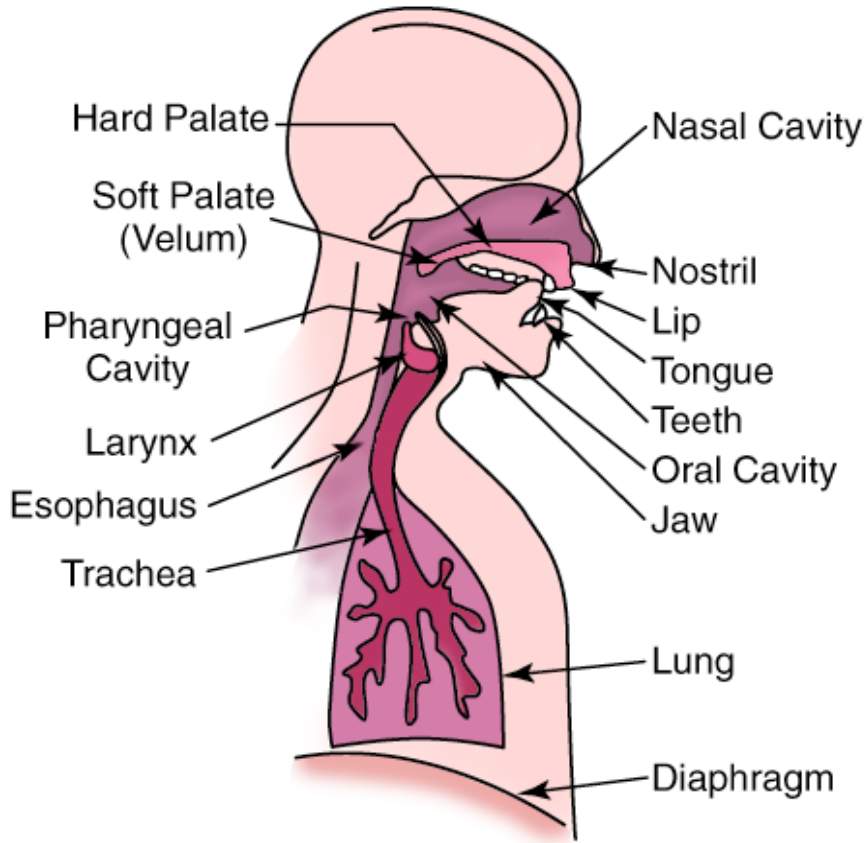

شکل ۳-۱- آناتومی گفتار انسان

### ۳-۴-۲- تاریخچه شنواسازی کامپیوترها

شاید اولین بار توماس ادیسون بود که توانست با اختراع گرامافون صدای انسان‌ها را ذخیره و پخش کند. این اولین تلاش انسان برای ضبط، پخش و پردازش صدا بود. سپس در سال‌های آینده این حرفه با اختراع ضبط صوت و دیگر سامانه‌ها گسترش یافت تا با ورود کامپیوترها به عرصه زندگی انسان‌ها این تلاش‌ها وارد حوزه باینری کامپیوترها شد و پردازش بر روی داده‌های صوتی به‌صورت رسمی آغاز گشت. [۳۱]

### ۳-۴-۳- تاریخچه تشخیص گفتار واج‌بنیان

در سال ۱۹۸۹ برای اولین بار تشخیص واج بنیان گفتار (تشخیص گفتار به صورت جزئی و در حد واج) معرفی شد. در آنجا با استفاده از شبکه عصبی پرسپترون و با استفاده از پردازش سیگنال ابتدایی، پژوهشگران توانسته بودند ۴ واج در زبان انگلیسی را شناسایی کنند. [۳۲]

در سال ۱۹۹۵ نیز مدل‌های زبان برای تبدیل گفتار به متن ارائه شدند؛ اما در سال‌های اخیر با پیشرفت کامپیوترها و افزایش سرعت آن‌ها، شاهد آن هستیم که در مرزهای دانش، پیشرفت‌های قابل‌توجهی در زمینه پردازش گفتار واج بنیان و همچنین ارائه مدل‌های زبان در زبان‌های مختلف برای تبدیل گفتار به متن ارائه می‌شوند. [۳۳]

## ۳-۵- نتیجه‌گیری

در این فصل به بررسی کارهای گذشته درزمینه‌ی پردازش و تشخیص اصوات گفتار توسط رایانه‌ها پرداخته شد. همان‌طور که مشاهده شد، پردازش سیگنال صوتی از ابتدایی‌ترین عملیات بر روی سیگنال‌های صوتی گفتار است تا جزئیات سیگنال استخراج گردد. در مرحله بعدی نحوه استفاده از الگوریتم‌های کلاس‌بندی به‌خصوص الگوریتم شبکه‌های عصبی مورداستفاده در کارهای گذشته، بیان گردید. در سال‌های اخیر مدل‌های زبان بر پایه تشخیص گفتار و تبدیل آن به متن نیز استفاده شده است تا مشکلات تشخیص گفتار بر پایه مدل زبان در زبان‌های مختلف کاهش یابد.

در فصل آینده به بررسی مراحل انجام پایان‌نامه و الگوریتم‌های مورد استفاده در آن پرداخته خواهد شد.

# فصل ۴:

# ارائه راه حل پیشنهادی

## ۴-۱- مقدمه

در این بخش به بررسی چالش‌های پیش رو در مبحث تشخیص اصوات به‌خصوص تشخیص گفتار خواهیم پرداخت و راه‌حل پیشنهادی خود را نیز مطرح خواهیم کرد.

## ۴-۲- تشخیص گفتار

### ۴-۲-۱- چالش‌ها

با توجه به مباحث مطرح‌شده در فصل گذشته، از مهم‌ترین چالش‌های مطرح در پردازش گفتار می‌توان به موارد زیر اشاره کرد:

۱- کوچک بودن و یا گاهی ظاهر نشدن واج در بیان گفتار
۲- تداخل سیگنال صوتی واج‌ها در حوزه زمانی یا فرکانسی
۳- عدم وجود قدرت کافی در پردازنده‌ها و الگوریتم‌ها برای تشخیص مناسب
۴- عدم وجود مدل کامل زبان در تشخیص کلمات و جملات
۵- به هم ریختن معنا و مفهوم کلی در صورت وقوع خطایی کوچک

### ۴-۲-۲- راه‌حل‌ها

در این مبحث، راه‌حلی ارائه خواهد شد تا تمامی چالش‌های گذشته را پوشش داده و تا حد امکان به رفع مشکلات حاصل از چالش‌ها بپردازد.

تشخیص گفتار در سیستم گفتاری و شنوایی انسان از خروج صدا از حنجره شروع شده و درنهایت به تشخیص در مغز ختم می‌شود. زمانی که صدا از حنجره در حال خروج است، ابتدا به‌صورت نفسی است که از ریه بیرون می‌آید. سپس با برخورد با تارهای صوتی به صدا تبدیل می‌شود. عمل تفاوت صوتی واج‌ها در حنجره اتفاق نمی‌افتد. در حنجره تنها شدت اصوات با یکدیگر بنا بر فشاری که نفس بر تارهای صوتی وارد می‌آورد مشخص می‌شود. زمانی که هوا به تارهای صوتی می‌رسند، تارهای صوتی گشوده هستند و زمانی که فرد می‌خواهد صدا انتشار دهد، تارهای صوتی به هم نزدیک می‌شوند و در فاصله کوتاهی از هم قرار می‌گیرند که باعث تولید صدا می‌شود. طول تارهای صوتی در مردان نازک‌تر و طویل‌تر از زنان است. طول این تارها

برای مردان در حدود ۱٫۷۵ سانتی‌متر تا ۲٫۵ سانتی‌متر است، درحالی‌که این طول برای زنان ۱٫۲۵ سانتیمتر تا ۱٫۷۵ سانتی‌متر است. درنتیجه صدای مردان بم و صدای زنان زیر است و این مشخص‌کننده بازه فرکانسی صدای مردان و زنان است. میزان ارتعاش تارهای صوتی بین ۱۱۲ تا ۲۴۰ ارتعاش در ثانیه است. [۳۴] تفاوت تلفظ واج‌ها در شکل قرار گرفتن زبان و لب‌ها و دندان‌ها صورت می‌گیرد. زمانی که صدا از عمق حنجره خارج می‌شود تنها یک صدای بی‌محتواست. این شکل دهان است که به آن محتوا می‌بخشد و آن را تبدیل به واج خاصی می‌کند. این تفاوت باعث می‌شود، تغییرات فرکانسی در بازه‌ی تولید صدا ایجاد شده و فرکانس‌های متفاوتی در بازه فرکانسی شخص صورت گیرد که باعث تولید واج‌ها شود. در تشخیص گفتار کامپیوتر هدف تشخیص این تغییرات در نمونه‌های فرکانسی است.

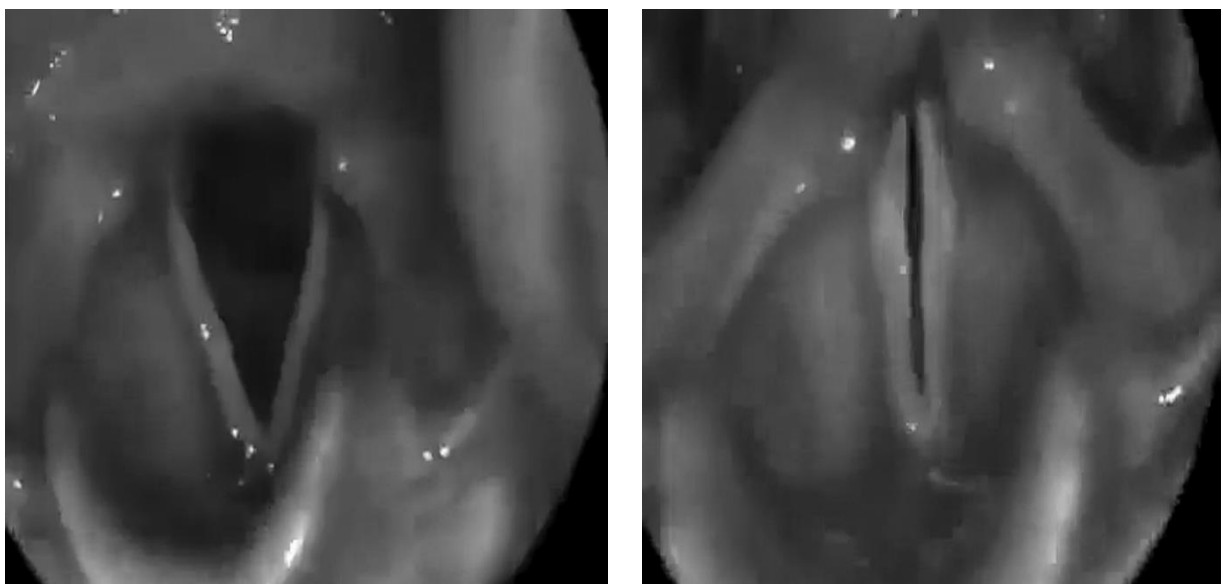

شکل ۴-۱- تارهای صوتی به حالت باز و نیمه باز

حال اگر تشخیص گفتار ماشینی همانند تشخیص گفتار انسانی شبیه‌سازی شود، مرحله تبدیل صدا به داده‌های رایانه‌ای، همان مرحله رسیدن اطلاعات شنوایی به مغز است و تا این بخش، تمامی مراحل سخت‌افزاری است؛ اما در مرحله اول تشخیص، شدت و فرکانس صدا در مغز تجزیه و تحلیل‌شده و بخش‌های مختلف آن از هم تفکیک می‌شود.

با توجه به این مسئله در راه‌حل پیشنهادی این پژوهش، گفتار تا حد کوچک‌ترین واحد زبانی (واج) تقسیم می‌شود. این عملیات باعث استخراج جزئی‌ترین ویژگی‌های ممکن از گفتار می‌شود. برای اینکه هم بتوان واج‌ها را تشخیص داد و هم در صورت مفقود شدن واج‌ها در گفتار مشکل چندانی برای تشخیص به وجود

نیاید، تشخیص واج‌ها را به تشخیص سیلاب‌ها تعمیم می‌دهیم. البته در این حالت باز تشخیص واج‌ها مورد هدف است، اما در میان تعمیم دادن واج به کلمه، از تعمیم واج به سیلاب استفاده می‌کنیم. سپس عملیات پردازش سیگنال جهت تشخیص شدت و فرکانس‌های موجود در صدا بررسی‌شده و فرکانس‌ها و شدت‌ها به ترتیب طبقه‌بندی شدند. در مرحله بعدی همانند عملیات مغز بر روی‌داده‌های گفتار، الگوریتم متمایزسازی و کلاس‌بندی بر روی‌داده‌های صوتی گفتار انجام می‌پذیرد. بدین‌صورت هرکدام از واج‌ها به‌صورت مجزا تشخیص داده‌شده و عملیات تشکیل سیلاب جهت تشخیص کلمات انجام می‌شود.

پس از تمایز بخش‌های مختلف داده صوتی و تشخیص آن‌ها، بحث تجربیات و آموخته‌های انسانی به میان آمده و از مدل زبان و پردازش زبان‌های طبیعی استفاده می‌شود که به طور کامل وابسته به زبان گوینده است. با استفاده از مدل زبان در صد تشخیص بالاتر می‌رود اما این به آن معنا نیست که عملیات تشخیص نمی‌تواند از کوچک‌ترین واحد زبانی (واج) آغاز شود. پشت سر هم قرار دادن این مراحل و همچنین کاستن از پیچیدگی آن‌ها در عین زیاد بودن تعداد مراحل می‌تواند در افزایش سرعت تشخیص نیز مؤثر باشد. درصورتی‌که از مدل زبان استفاده کنیم می‌توانیم حدس بزنیم که سیلاب تشخیص داده‌شده مخصوص کدام کلمه بوده و همچنین تشخیص می‌دهیم که کلمه در کجای جمله جای داشته است و اینکه با چه احتمالی ممکن است در چنین جمله‌ای ظاهر شود.

## ۴-۳- مراحل انجام طرح

در ابتدا که تصمیم به شروع برای پردازش گفتار گرفته شد، بررسی‌ها برای یافتن ایده‌ای کم‌نظیر برای بهبود شرایط آن در این زمینه آغاز شد. زمینه‌های متفاوتی همچون تشخیص گوینده، تشخیص دستگاه‌های صوتی و تشخیص گفتار مطرح بود. در میانه تحقیقات ایده‌ای به ذهن رسید که در زبان فارسی بی‌سابقه بود. تشخیص گفتار با رویکرد واج بنیان هرچند توسط برخی پژوهشگاه‌های مطرح دنیا در حال بررسی و پیشرفت است اما در زبان فارسی شواهدی در دست است که هنوز به انجام نرسیده است. علت مشخص نبودن نتایج این پژوهش‌ها به دلیل اهمیت بالای این پژوهش است. ازآنجایی‌که برخی پژوهش‌ها، همانند پژوهش درزمینه موتورهای جستجو، در بهبود شرایط شرکت‌ها اهمیت بالایی دارند، نتایج این پژوهش‌ها تنها به‌صورت خروجی اجرایی منتشر می‌شود. از همین رو پژوهش‌های انجام شده در این زمینه تنها استفاده از الگوریتم‌های بسیار اولیه برای ساخت الگوریتم‌های کاربردی درزمینه‌ی تشخیص گفتار هستند. شرایط در این راستا برای شروع پژوهش بسیار سخت بود اما با توجه به الگوریتم‌های موجود در این زمینه برای ساخت برنامه‌های کاربردی

مناسب، نتیجه ممکن می‌نمود. در ابتدا پژوهش از الگوریتم‌های پایه‌ای بسیار ابتدایی شروع شد. ابتدا سه نمونه صوتی شامل تلفظ تمام دو واجی‌های ممکن شامل مصوت و صامت، توسط برنامه طراحی شده در نرم‌افزار متلب تهیه گردید تا به صورت مداوم اقدام به ضبط اصوات و نمایش نمودار شدت-زمان آن‌ها کند. در ظاهر، این نمودارها تنها، اطلاعات موجود حاصل از شدت صدا و زمان اتفاق افتادن آن بود. اگر این نمودارها در باطن نیز تنها اطلاعاتی از شدت صدا و زمان بود باید نمودار آن به‌صورت خطی صاف با بالا و پایین شدن‌های حاصل از کم‌وزیاد شدن در واحد زمان بود؛ اما در نمودار شدت-زمان، نمودار به‌صورت متناوب است که علاوه بر شدت و زمان واحد دیگری به نام طول‌موج نیز به چشم می‌خورد. به هر مقدار که طول‌موج‌ها کمتر با شد فرکانس بالاتر است؛ بنابراین در اصل این نمودارها دارای هر سه ویژگی زمان، شدت و فرکانس است و ویژگی‌های لازم برای تشخیص گفتار و تشخیص صوت، همین سه ویژگی هستند. البته یافتن این اطلاعات در نمودار اولیه صوتی که معروف به نمودار حوزه زمانی است، بسیار دشوار است. برای مثال هرچند اطلاعات مربوط به طول‌موج در حوزه زمانی وجود دارد ولی یافتن این مسئله که چه فرکانس‌هایی در صدا موجود است پیچیده بوده، نیاز به محاسبات فراوانی دارد و نمی‌توان از روی محورهای مختصاتی نمودار، مقدار آن‌ها را مشاهده نمود؛ اما سؤال اینجاست که این محاسبات چه الگوریتم‌هایی دارند. یکی از محبوب‌ترین این الگوریتم‌ها، الگوریتم تبدیل فوریه بود.

## ۴-۳-۱- تبدیل فوریه

در مرحله دوم، تبدیل فوریه بر روی نمودار حوزه زمانی اعمال شد تا به حوزه فرکانسی تبدیل گردد. پس از به دست آمدن نمودارهای حوزه فرکانسی و مشاهده مقادیر فرکانسی سعی شد تا تفاوت‌های نمودارها برای کلاس‌بندی مشخص گردد. ازآنجایی‌که در این مرحله سعی می‌شد تا تفاوت‌ها به‌صورت دیداری بررسی شوند آزمایش‌ها با نمونه‌گیری آنی انجام می‌شد. در مراحل اولیه برخی تفاوت‌ها به‌صورت دیداری مشاهده می‌شد اما با توجه به جزئیات بسیار ریز در موارد آزمایش تفاوت محسوس نبود و همچنین این امکان وجود نداشت که به این تفاوت‌های ریز اعتماد کرد. از همین رو تصمیم بر آن شد که تعداد نمونه‌ها افزایش یابد تا این نمونه‌ها بیشتر و بهتر مشاهده شود.

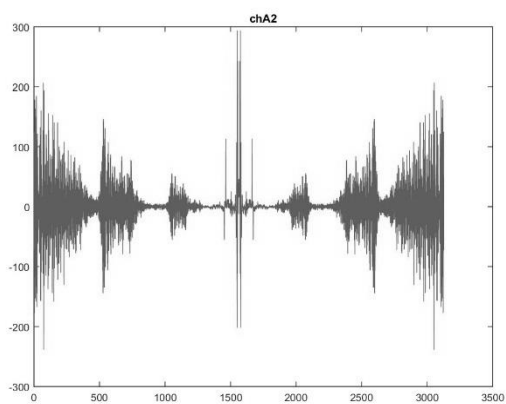

شکل ۲-۴- نمونه‌ای از تبدیل فوریه از ترکیب دو واج "چ" و "آ"

### ۴-۳-۲- ایجاد نمونه‌ها

در مرحله سوم تصمیم به ایجاد نمونه‌های آزمایشی گرفته شد. در این مرحله تصمیم‌گیری برای شکل‌دهی به نمونه‌های آزمایشی که بتوان در تمام مراحل بعدی نیز بتوان بدون تغییر از آن استفاده کرد بسیار سخت می‌نمود. چراکه رویکرد واج بنیان در زبان فارسی بی‌سابقه بوده و نمونه‌های دیگر زبان‌ها نیز بدون وجود اطلاعات و پایگاه‌های داده‌ای در حد واج بود. در این مرحله چون واج‌های صامت بدون وجود واج‌های مصوت قابل تلفظ نبودند وجود واج‌های مصوت در کنار واج‌های صامت لازم می‌نمود. استفاده از رویکرد واج بنیان در تشخیص گفتار هم‌اکنون توسط شرکت‌های بزرگ پژوهشی در این زمینه آغاز شده است و بهترین روش ترکیب حداقل ترکیب‌های واج است که بتوان تلفظ کرد. همچنین درزمینه تشخیص منفرد واج‌های صامت زبان چک را می‌توان مثال زد. حتی در زبان چک که حروف صامت پشت سرهم تکرار می‌شوند، باز مصوتی در میان صامت‌ها حاضر می‌شوند. یکی از واضح‌ترین مثال‌هایی که می‌توان در این مورد زد مثال کلمه "کارد" در زبان فارسی است. در این کلمه حروف "ر" و" د " پشت سرهم ظاهرشده‌اند. در تلفظ فارسی به این دلیل که این کلمه تنها شامل دو واج صامت پشت سر هم است، وجود واجی صامت در میان این دو واج احساس نمی‌شود اما درصورتی‌که صدای گوینده را ضبط کرده و سرعت تلفظ را کم کنیم خواهیم دید که واجی تقریباً مشابه واج " شعاع " در زبان انگلیسی در میان این دو واج وجود دارد. با در نظر گرفتن این مسائل و کنار گذاشتن استثنائاتی نادر مانند کلمه" کارد " یا " کادر " در فارسی تصمیم بر آن شد تا با ترکیب واج‌های مصوت و صامت نمونه‌ها دارای دو واج به ترتیب صامت و مصوت باشند به‌طوری‌که مصوت‌ها تنها شامل شش مصوت اصلی فارسی (سه مصوت کوتاه و سه مصوت بلند باشند.) و از مصوت" شعاع" در آن استفاده نشود. صامت‌ها نیز با در نظر گرفتن این مورد که در فارسی هر چهار حرف" ز "،" ذ "،" ض " و" ظ " دارای تلفظ یکسانی هستند، تعداد صامت‌ها به تعداد حروف فارسی به‌جز" الف " نبوده و کمتر از این مقدار بود. این تعداد

در زبان فارسی برابر ۲۳ واج است. درواقع اگر الفبای فارسی به‌جای استفاده از الفبای عربی از الفبایی استفاده می‌کرد که همان‌طور که نوشته می‌شود خوانده شود، تعداد واج‌ها و درنتیجه حروف فارسی برابر ۲۳ بعلاوه ۶ می‌شد و از همین رو ترکیب عبارات آموزشی برابر ۱۳۸ نمونه می‌شد.

### ۴-۳-۳- تبدیل فوریه روی نمونه‌ها

در مرحله بعد تبدیل فوریه بر روی نمونه‌های آزمایشی مورد آزمایش قرار گرفت. پس از مشاهده تفاوت‌های تبدیل فوریه بر روی نمونه‌ها بیشتر تصمیم بر آن شد که در کنار تبدیل فوریه، تبدیل موجک نیز بر روی نمونه‌ها اعمال شود و تفاوت‌های این تبدیلات و همچنین تفاوت این تبدیلات با ترکیب این دو تبدیل سنجیده شود. پس از اعمال تبدیل موجک و همچنین هر دو تبدیل فوریه و موجک بر روی نمونه‌های صدا، عمل مقایسه دوباره بر روی نمونه‌ها صورت گرفت. در این مقایسه شباهت‌هایی در برخی از نمودارهای فوریه و موجک دیده می شد که ابتدا حدس بر این بود که این شباهت‌ها همان شباهت‌های مدنظر برای تشخیص واج‌های یکسان است، اما با آزمایش‌ها و انطباق‌هایی که بین نمودارها و نمونه‌ها صورت گرفت، مشاهده شد که این شباهت‌ها میان نمودارهای برخی واج‌های مصوت مشابه، به خاطر نوع صدای گوینده بوده و این شباهت‌ها درواقع شباهت میان نمودارهای تلفظ بلند واج مصوت توسط یک گوینده است.

در مرحله بعدی تصمیم بر آن شد که با تحلیل آماری نمونه‌ها تفاوت‌ها و میزان تفاوت‌ها در نمونه‌های مختلف با یکدیگر سنجیده شود تا میزان تفاوت‌ها در نمونه‌های مختلف از هم تفکیک شوند. بدین‌وسیله امکان تشخیص واج‌های مشابه و تمایز واج‌های متفاوت تا حدودی مشخص می‌شد.

### ۴-۳-۴- شبکه عصبی مصنوعی

با توجه به اینکه تحلیل‌های آماری دقت مناسب جهت تشخیص نمونه‌ها را ارائه ندادند نیاز برای استفاده از الگوریتمی نوین جهت تشخیص هرچه بهتر واج‌ها از روی نمونه‌های صوتی بیش‌ازپیش حس می شد. در این زمینه روش‌های مختلفی وجود داشت. یکی از این روش‌ها، روش ماشین بردار پشتیبان (SVM[1]) بود. در این روش هر داده‌ای به صورت برداری وارد ماشین بردار پشتیبان شده و در دسته‌بندی‌های متمایز داده قرار می‌شدند. برای تشخیص تفاوت‌های میان نمونه‌ها زمانی که از ماشین بردار پشتیبان استفاده می‌شد، به دلیل ساختار برداری و تک‌بعدی، برخی از ویژگی‌های صوتی کمرنگ می شد. همچنین به دلیل استفاده از روش

---

[1] Support Vector Machine

تعیین مقدار ثابت برای محدوده کلاس‌ها این الگوریتم توانایی تشخیص با دقت بالا برای نمونه‌های حساس، پیچیده و متراکمی مانند نمونه‌های صوتی را نداشت؛ بنابراین پس از تحقیق بر روی الگوریتم‌های کلاس‌بندی و بررسی ویژگی‌های نمونه‌های صوتی مشاهده شد که سیگنال‌های صوتی بسیار بیشتر از دیگر سیگنال‌ها تحت تأثیر نوع مرزبندی کلاس‌ها قرار می‌گیرد. برای مثال در الگوریتم تشخیص SVM پس از آموزش ماشین بردار پشتیبان، محدوده کلاس‌ها با مرزبندی ثابت از یکدیگر جدا می‌شدند، درحالی‌که در الگوریتمی مانند شبکه عصبی، کلاس‌ها باهم همپوشانی داشتند و تنها درصد تشابه نمونه آزمایشی باکلاس‌ها بود که میزان تطابق نمونه آزمایشی با هرکدام از کلاس‌ها را مشخص می‌کرد.[۳۵] با توجه به وجود مؤلفه‌های متنوع در صدا مانند فرکانس، شدت و زمان، دسته‌بندی این کلاس‌ها به‌صورت همپوشانی و درصدی پاسخگوی بهتری برای کلاس‌بندی داده‌ها بود. سپس برای تست اولیه از شبکه عصبی مصنوعی پرسپترون با یک‌لایه پنهان از نوع تشخیص الگو در نرم‌افزار متلب استفاده شد. برای این منظور باید داده‌هایی از تصاویر نمودارهای صوتی به شبکه داده می‌شد تا شبکه می‌توانست داده‌ها را پردازش کرده و نتیجه نهایی را به در کلاس‌های مربوط به واج‌های مختلف ارائه دهد.

## ۴-۳-۵- تست نمودارها بر روی شبکه عصبی

همان‌طور که در بخش‌های گذشته نیز ذکر شد، در مرحله اول ورودی شبکه نمودارهای تبدیل فوریه، تبدیل موجک و ترکیب این دو نمودار بودند. با توجه به فاصله زیاد آوایی واج‌های مصوت نسبت به یکدیگر ابتدا نمونه‌ها در ۶ کلاس با مصوت‌های متفاوت دسته‌بندی‌شده و به شبکه عصبی مصنوعی آموزش داده شدند. برای این منظور نمودارهای تبدیل فوریه آن‌ها به‌عنوان ورودی به شبکه عصبی داده شد. پس از آموزش نمونه‌ها و ذخیره شبکه آموزش داده‌شده، نمونه‌هایی برای تشخیص به‌صورت زنده ضبط‌شده و به برنامه جهت تشخیص داده شد. در این آزمایش درصورتی‌که فاصله کاربر با سیستم رعایت می‌شد تا مشکلات سخت‌افزاری مانندِ نویز حاصل از ضبط صدا، هماهنگ با نمونه‌های آموزشی بود. نمونه‌های آزمایشی با احتمال ۹۵ درصد تشخیص داده می‌شدند. با در نظر گرفتن این موضوع که تشخیص ۹۵ درصدی واج‌ها با استفاده از ساخت کلمه با لغت‌نامه می‌توانست درصد بسیار بالایی برای تشخیص کلمه و در راستای آن درصد بسیار بالاتری برای تشخیص جمله را در پی داشته باشد. این درصد تشخیص بسیار ایده‌آل بود؛ اما در این میان دو مشکل وجود داشت. مشکل اول تشخیص واج‌های صامت بود که بسیار دشوارتر از تشخیص واج‌های مصوت بود و مشکل دوم نمودارهای ورودی شبکه بودند. این نمودارها که از نوع تبدیل فوریه و موجک بودند، نمی‌توانستند تمامی پارامترهای موجود در صدا را برجسته کرده و باعث تشخیص ساده‌تر توسط شبکه عصبی شوند. عدم

اطلاعات کافی در این نمودارها درواقع مهم‌ترین عامل ایجاد مشکل اول نیز بود. از همین رو انجام پیش‌پردازش بر روی نمونه‌های صدا برای ورود به شبکه عصبی لازم می‌نمود و همین مسئله که شبکه عصبی نمی‌توانست به‌خودی‌خود اطلاعات موجود در نمودارهای مختلف را استخراج کند، نشان از پیچیدگی نمونه‌های سیگنال‌های صدا نسبت به سایر انواع سیگنال‌ها بود.

## ٤-٣-٦- افزایش دقت مصوت‌ها

پس از عدم دریافت دقت مناسب از نمودارهای تبدیل فوریه و موجک به‌عنوان ورودی شبکه عصبی، تحقیقات برای یافتن نمونه‌های مناسب‌تری از الگوریتم‌های استخراج ویژگی آغاز شد. در این میان الگوریتم‌های مختلفی همانند MFCC و STFT مطرح بودند. برای شروع نیاز به استفاده از الگوریتم MFCC بهینه و آماده‌ای بود تا بر روی داده‌های آموزشی مورداستفاده قرار گیرد. برای همین منظور الگوریتم MFCC برای استخراج ویژگی‌های حوزه فرکانسی صدا که با تبدیل صوت به تصویر امکان استخراج ویژگی‌های فرکانسی را به صورت تصویر فراهم می‌کرد، دریافت شد. این الگوریتم برای تشخیص اصوات محیطی مانند اصوات ترافیک شهری استفاده می‌شد. از همین رو این الگوریتم به‌گونه‌ای برای تشخیص اصوات محیطی با طول زمانی بلند و تفاوت فرکانسی و گاهی شدت زیاد بهینه‌شده بود. همچنین الگوریتم MFCC به دلیل استخراج ویژگی‌های بسیار جزئی حوزه فرکانسی قادر به تشخیص مناسب اصوات طولانی ازنظر زمانی و همچنین پر جزئیاتی مانندِ تفاوت میان واج‌های مصوت نبود. از همین رو در اولین آزمایش الگوریتم MFCC بهینه شده توسط دانشگاه لوون[1]، بر روی نمونه‌های مصوت، با شکست مواجه شد. پس‌ازاین شکست تنها الگوریتم باقی‌مانده از سری الگوریتم‌های تشخیص اصوات در حوزه فرکانسی و زمان-فرکانس، الگوریتم STFT بود. این الگوریتم درواقع تمامی اتفاقات موجود در صدا را در یک نمودار سه‌بعدی زمان، فرکانس و شدت نشان می‌داد. برخلاف MFCC در STFT تمامی اتفاقات با مقادیر کاملاً مشخص بر اساس برچسب‌های نمودارها و به صورت کاملاً مرتب بود. پس از استخراج ویژگی‌های صدا توسط این الگوریتم، نمودارها برای تشخیص به شبکه عصبی واگذار شدند. پس از انجام تمرین و آموزش توسط این نمودارها، نتایج به‌دست‌آمده نشان از بهبود محسوس نتایج به‌دست‌آمده برای تشخیص واج‌های مصوت بود که باعث ایجاد امیدواری برای تشخیص واج‌های صامت توسط این الگوریتم می‌شد.

---

[1] Luven

### ۴-۳-۷- بررسی بهینگی تشخیص واج‌بنیان

در مرحله بعدی که واج‌های مصوت توسط الگوریتم STFT با دقت قابل قبولی تشخیص داده شدند، نحوه ارتباط واج‌های صامت و مصوت و همچنین نحوه تشخیص این واج‌ها کنار هم باید موردبررسی مجدد قرار می‌گرفت. در این مرحله، دریافتیم که تشخیص واج بنیان هم‌اکنون در دستور کار شرکت‌های بزرگ سرمایه‌گذار بر روی تشخیص گفتار قرارگرفته است. همچنین تشخیص واج بنیان در مراکز تحقیقاتی معمولاً با رویکرد خطی به صورت ساخت متوالی کلمات و جملات از واج‌ها صورت می‌گیرد. در رویکرد واج بنیان با استفاده از رویکرد خطی، معمولاً ابتدا واج‌ها تک‌به‌تک موردبررسی و پیش‌پردازش و تشخیص قرار می‌گیرند و سپس همین عمل پیش‌پردازش و تشخیص بر روی دو واجی‌ها و سپس سه واجی‌ها و... صورت می‌گیرد تا زمانی که بتوان ساختاری به صورت کلمه و سپس جمله به وجود آورد. هرچند در مراکز تحقیقاتی هم‌اکنون اولویت بر ساخت کل ساختار جمله با استفاده از الگوریتم‌های کلاس‌بندی و تشخیص است؛ اما در رویکرد انتخابی در این پژوهش، از رویکرد تشخیص تک واجی و سپس ساخت کلمات با همین ساختار تک واجی استفاده شد. به این صورت که ابتدا تمامی واج‌ها تک‌به‌تک مورد پردازش صوت و تشخیص واج قرار می‌گرفتند و سپس تمامی گزینه‌های ممکن مورد بررسی قرار می‌گرفت. این رویکرد از جهت اهمیت تشخیص واج‌های منفرد کاری بسیار دشوارتر از موردبررسی قرار دادن ساختارهای چند واجی بود، اما این همان رویکرد موردا ستفاده در مغز و ساختار شنوایی انسان بود که بعدها به دلیل سختی تشخیص تبدیل به تشخیص سیلاب‌ها و کلمات شده بود. درواقع رویکرد این پژوهش بر این اساس بود که اعلام کند بازگشت به ساختار آفرینش شنوایی هوش انسان در هوش مصنوعی عملی دور از واقعیت نیست. این رویکرد هرچند سخت ولی ممکن است و استفاده از این رویکرد می‌تواند ما را به دقت انسان برای تشخیص گفتار و حتی بسیار بالاتر از این دقت برساند. دقتی که سال‌هاست شرکت‌های بزرگ فناوری طبق اعلام خودشان در تبدیل آن به واقعیت به مشکل خورده‌اند و حتی دست یافتن به‌دقت تشخیص انسان که حاصل از خطاهای فراوان انسانی است نیز ممکن نیست.

### ۴-۳-۸- ویژگی‌های STFT

روش STFT یا همان تبدیل فوریه با بازه زمانی کوتاه، درواقع تبدیل فوریه‌ای است که استخراج فرکانس‌های صدا در بازه‌های کوتاه زمانی صورت می‌گیرد. در تبدیل فوریه معمولی، عمل تبدیل در کل بازه زمانی رخ دادن صدا صورت می‌گیرد و این مسئله باعث می‌شود تا فرکانس‌های ظاهرشده در بازه مشخصی از

صدا را در دست نداشته باشیم؛ اما در تبدیل فوریه زمان کوتاه، این امکان فراهم می‌شود که در بازه‌های زمانی کوتاه و دلخواه شدت صدای رخ دادن هر فرکانس را داشته باشیم.

باوجود مشکلات سخت‌افزاری و نرم‌افزاری مانند وجود نویزهای سنگین در صدا یا عدم تطبیق صحیح کانال‌های صدا و همچنین تفاوت صدای افراد در تلفظ واج‌های مختلف، نیاز بود که خطای تشخیص STFT هرچه بیشتر پایین آمده و همچنین کاملاً غیر وابسته به گوینده یا کانال‌های صدا یا موقعیت نویز با شد. با توجه به این موضوع که صدای گوینده‌های مختلف چه مرد، چه زن و چه کودک و حتی در همین سه دسته، تمام دسته‌های مختلف و انواع صداها، دارای موقعیت فرکانسی مختلف در واج‌های مختلف بودند، اگر بررسی و تشخیص نمونه‌های نمودار STFT به صورت مقایسه آماری صورت می‌گرفت، قطعاً تمامی نمونه‌ها نسبت به یکدیگر کاملاً متمایز بودند. همچنین در صورت بررسی نمونه‌ها با الگوریتم‌های کلاس‌بندی خطی مانند SVM یافتن الگوها در بردارهای خطی بسیار مشکل و کم‌دقت بود. همچنین با توجه به ساختار سایر الگوریتم‌های کلاس‌بندی که داده‌ها را در دسته‌های کاملاً مجزا از هم دسته‌بندی می‌کردند امکان تشابه سنجی دقیق این نمونه‌ها برای تشابه سنجی بخش‌های مختلف صدا وجود نداشت.

درروش STFT با توجه به وابستگی به زمان، شدت و فرکانس و همچنین توجه به این موضوع که مصوت‌ها در صدای شخصی خاص به صورت کشیده و در یک بازه فرکانسی خاص با تغییرات کوتاه ظاهر می‌شوند، صداهای مصوت بر روی نمودار STFT به صورت خطوطی تقریباً صاف با خمیدگی‌هایی در ابتدا و انتهای آنان که حاصل از تغییرات شدت و فرکانسی صدا در بازه زمانی بود ظاهر می‌شدند. درواقع برای تشخیص صدا از طریق نمودارهای STFT، همین نمودارها بودند که نقش اصلی را ایفا می‌کردند. در اصطلاح پردازش صوت به این خطوط در نمودار STFT، Formant می‌گویند. فرمنت‌ها یکی از مهم‌ترین معیارهای تشخیص اصوات بلند

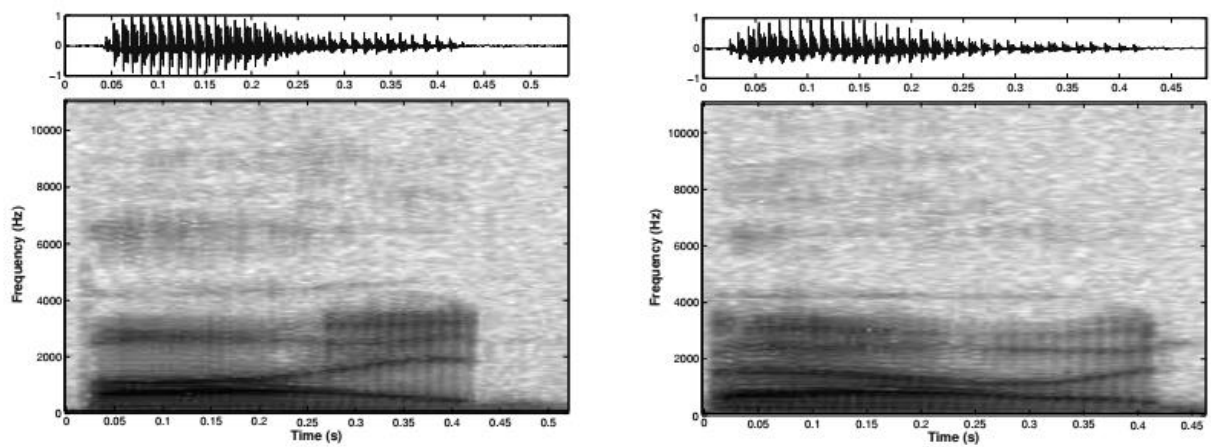

شکل ۴-۳- تفاوت فرمنت‌های دو کلمه bide و bowed

مانند اصوات مصوت در تشخیص اصوات هستند. محل قرار گرفتن این فرمنت‌ها و همچنین میزان خمیدگی و تعداد این فرمنت‌ها مشخص‌کننده نوع صدای بلند است.

با برش نمودار و استخراج بخش‌های مهم نمودار (مانند بخش‌های پررنگ مربوط به صداهای مصوت) و همچنین تبدیل نمودارها به نمودارهای سیاه‌وسفید برای افزایش سرعت و دقت پردازش تصاویر، درصد تشخیص صداهای مصوت، با در نظر گرفتن خطای سامانه‌ای و سخت‌افزاری تقریباً دقیق بود و در تمامی موارد عمل تشخیص به‌درستی صورت می‌گرفت.

با توجه به ویژگی‌های الگوریتم تبدیل فوریه زمان کوتاه استفاده از این الگوریتم برای استخراج ویژگی‌های ریز فرکانسی صداهای صامت بسیار سخت می‌نمود. از همین رو دوباره تلاش‌ها برای تشخیص مناسب صداهای صامت آغاز شد. برخی مقالات بر روی تشخیص منفرد صامت‌ها کارکرده‌اند؛ اما هنوز صامت‌ها در کنار مصوت‌ها بهتر تشخیص داده می‌شوند. بهترین روش برای تشخیص صامت‌ها، آموزش سیستم خود بر روی نمونه‌های زیاد و متفاوتی از ترکیب‌های صامت‌ها است. البته در هیچ زبانی صامت‌ها به‌تنهایی ظاهر نمی‌شوند. حتی در زبان چک که حروف صامت بیشتر در کنار هم ظاهر می‌شوند، یک مصوت در میان آن‌ها وجود دارد. تشخیص صامت‌ها توسط آموزش سیستم بر روی هزاران کلمه بود که صامت‌ها نیز در داخل کلمه شناسایی می‌شدند؛ اما با توجه به رویکرد واج بنیان این پژوهش باید روشی استفاده می‌شد که واج‌ها را بدون در نظر گرفتن واج‌های کناری در عبارات تشخیص دهد.

### ۴-۳-۹-۴- بازگشت دوباره به روش MFCC

پس از بررسی‌ها بر روی الگوریتم‌های متفاوت، مشاهده شد که در اکثر پژوهش‌های قبلی، از ابتدا تاکنون، روش MFCC پرتکرارترین روش برای تشخیص صامت‌ها بوده است. هرچند در روش‌های نوین تکیه بر یادگیری ژرف و روش‌های تشخیص داده‌های خام بیشتر مورداستفاده قرارگرفته بودند و تکیه کمتری بر الگوریتم MFCC شده بود، اما این الگوریتم هنوز به‌طور گسترده مورداستفاده قرار می‌گرفت. با توجه به شکست استفاده از این الگوریتم در مراحل ابتدایی در استفاده دوباره از این الگوریتم شک وجود داشت اما با توجه به نتایج سایر پژوهش‌ها دوباره مورد استفاده قرار گرفت. این الگوریتم زمانی که در مرحله اول با نسخه پیاده‌سازی شده توسط دانشگاه لوون مورداستفاده قرار گرفت، اما با توجه به وظیفه متفاوت این الگوریتم در این پژوهش، در مرحله دوم پیاده‌سازی دیگری مورداستفاده قرار گرفت. این پیاده‌سازی درواقع برخلاف مرحله اول تمام بازه‌های فرکانسی را بدون محدودیت استخراج می‌کرد و جزئیات بیشتری را در اختیار می‌گذاشت. قبل از استفاده از این الگوریتم نیز پیش‌پردازش‌هایی بر روی داده‌های صوتی انجام شد. با توجه

به اینکه نیاز بود داده‌های ا صوات  صامت دقیق‌تر برر سی  شوند، ابتدا الگوریتم حذف نویز بر روی این داده‌ها صورت گرفت؛ اما مشکلی در این میان وجود داشت. قبل از شروع اصوات صامت نویز وجود داشت و اصوات صامت در زمان شروع تلفظ بسیار شبیه به نویز بودند. برای مثال در تلفظ واج "س" شباهت بسیاری به نویز اطراف وجود دا شت دا شت چراکه نویز مانند تلفظ مداوم واج "س" بود. برای این منظور از ویژگی تارهای  صوتی و دستگاه شنوایی انسان برای تشخیص واج‌ها استفاده شد. در سیستم شنوایی انسان، صدای نویز نیست که از صدای نویز بالاتر باشد. برای مثال فرض کنید در کارخانه‌ای که دستگاه‌های پرسروصدا دارند در حال صحبت کردن هستید. هرقدر که صدای شما تفاوت فرکانسی بیشتری نسبت به صدای د ستگاه‌ها داشته با شد، باز صدای شما شنیده نخواهد شد؛ بنابراین مجبور خواهید بود صدایی بلندتر از صدای د ستگاه‌ها را در گوش مخاطب خود القا کنید. این تفاوت شدت صدا هر میزان هم که جزئی با شد، می‌تواند صدای شما را به گوش مخاطبتان برساند؛ بنابراین برای حذف نویز از تفاوت شدت صدای نویز و صدای صامت استفاده شد. با توجه به اینکه این تفاوت در صداهای مختلف متفاوت بود الگوریتم پویایی برای تشخیص این تفاوت طراحی‌شده و نویز حذف شد.

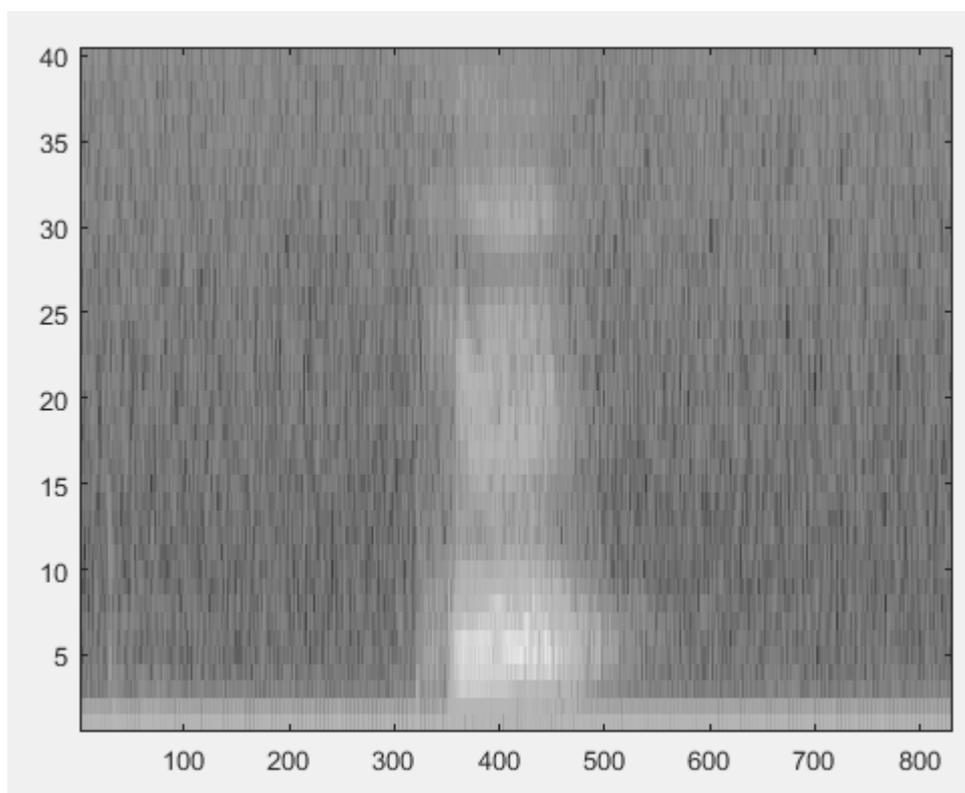

شکل ۴-۴- نمودار MFCC تلفظ دو واج "پ" و "آ"

در مرحله بعدی صدای مصوت بعد از صامت کوتاه شد تا صدای صامت برجستگی بیشتری داشته باشد و در مرحله آخر صدا به‌صورت ورودی وارد سیستم تشخیص شد. ابتدا صدا توسط الگوریتم MFCC تحلیل‌شده و نتایج به صورت تصویری وارد شبکه عصبی پردازش الگو شد. پس از آموزش سیستم با نتایج تصویری الگوریتم MFCC، مرحله نهایی فرارسید.

در این مرحله نتایج الگوریتم MFCC پس از آموزش سیستم برای تشخیص در شبکه عصبی ذخیره می‌شدند تا در مراحل بعدی مورد آزمایش قرار گیرند.

## ۴-۴- آزمایش نمونه‌ها

در مراحل قبلی توضیح داده شد که با استفاده از الگوریتم STFT، تشخیص نمونه‌های مصوت، به‌صورت کامل ممکن شد. در مرحله نهایی نمونه‌ها به شبکه عصبی داده شدند و درنتیجه شبکه عصبی به میزان قابل قبولی توانست نمونه‌ها را تشخیص دهید. در بیش از ۶۰ در صد نمونه‌ها، واج صامت به‌عنوان اولین گزینه‌ی پیشنهادی شبکه عصبی و در سایر موارد نیز جزو اولین گزینه‌های تشخیص داده‌شده توسط شبکه عصبی بود.

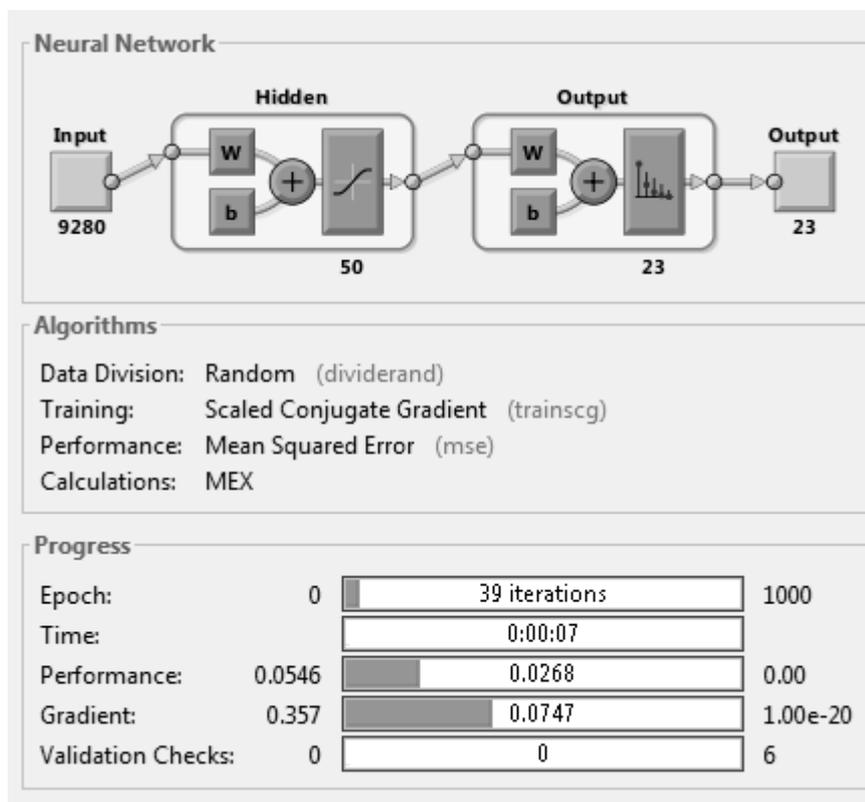

شکل ۴-۵- شبکه عصبی تشخیص اصوات واج‌ها

با در نظر گرفتن استفاده از پردازش متن برای تشخیص بهتر کلمات و جملات، میزان تشخیص به‌اندازه‌ای بالا بود که با در نظر گرفتن بهبود شرایط همانند شرایط سخت‌افزاری و نرم‌افزاری، امکان برابری تشخیص با دقت تشخیص انسان وجود داشت. هرچند هدف این پژوهش تشخیص عبارات گفتاری به‌طور قاطع و بسیار بالاتر از میزان تشخیص انسان بود، اما در ابتدای پژوهش رسیدن به میزان تشخیص انسان نیز هدفی دست نیافتی به نظر می‌رسید.

## ٤-٥- حذف نویز

در این مرحله با توجه به کافی نبودن در صد تشخیص واجهای صامت بر روی نمونه‌های گفتاری، نیاز به استفاده از شبکه‌های عصبی ژرف بیش‌ازپیش حس شد. همچنین به دلیل بهبود عملکرد سامانه‌های تشخیص گفتار و همچنین شبکه‌های عصبی مصنوعی ژرف بر روی تشخیص واجهای صامت، عملیات حذف نویز بر روی داده‌های گفتاری صورت گرفت.

برای این منظور از نرم‌افزار Adobe Audtion و حذف نویز به‌وسیله الگوریتم Adaptive Noise Reduction استفاده شد. در این روش توسط نرم‌افزار ابتدا محلی که دارای کمترین نمونه گفتاری و بیشترین نویز بود انتخاب شده و به‌عنوان اثرانگشت نویز به نرم‌افزار معرفی شد. سپس توسط الگوریتم مذکور حذف نویز بر روی کل نمونه صدا بر اساس نویز معرفی شده صورت گرفته و با تکرار این عمل نویز موجود تا حد بسیار زیادی حذف شد که کار را برای استفاده از الگوریتم STFT و همچنین تشخیص را برای شبکه عصبی آسان‌تر می‌کرد.

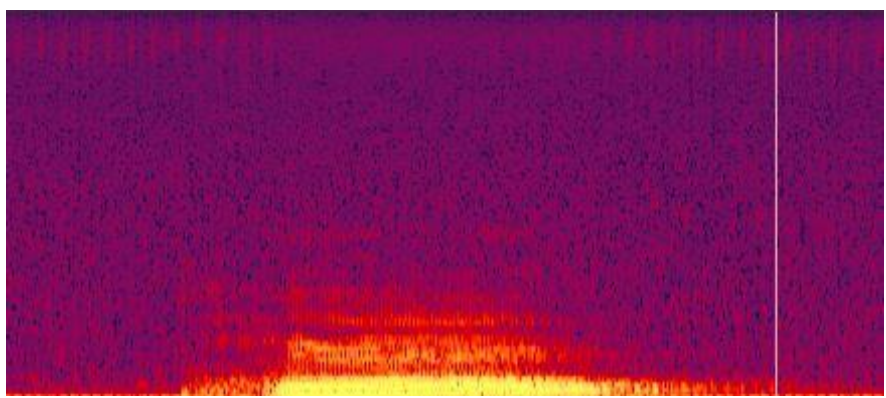

شکل ۴-۶- نمونه STFT گفتار قبل از حذف نویز

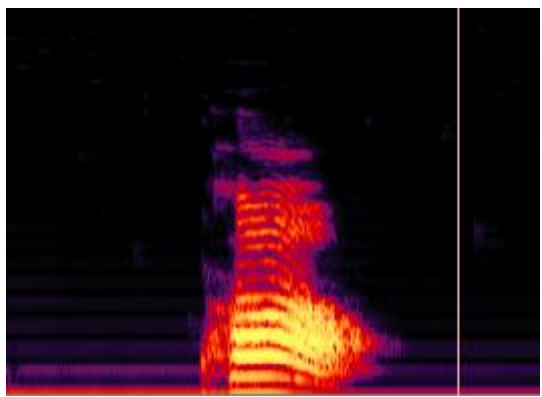

**شکل ۴-۷- نمونه STFT گفتار پس از حذف نویز**

## ۴-۶- شبکه عصبی مصنوعی ژرف

قبل از پرداختن به مبحث یادگیری ژرف و نحوه تشخیص واج‌ها در این معماری و فن، ابتدا واج‌های موجود در مجموعه داده می‌شوند.

| Persian Example | English form | Persian form |
|---|---|---|
| آل | A | آ |
| ایل | I | ای |
| او | ʊ | او |
| اول | æ | اَ |
| اسم | e | اِ |
| اردو | o | اُ |
| پا | P | پ |
| با | B | ب |
| تا | T | ت |
| دارو | D | د |
| چاقو | tʃ | چ |
| جارو | dʒ | ج |
| کاری | K | ک |
| گاری | G | گ |
| فاطمه | F | ف |
| واهمه | V | و |
| خاطره | Kh | خ |
| ساز | S | س |
| ز | Z | زار |
| ش | ʃ | شار |
| ژ | ʒ | ژاکت |
| م | M | ماکت |
| ن | N | نادی |
| ه | H | هادی |
| ل | L | لابه |
| ر | R | راهبه |
| ق | Q | قاری |
| ی | j | یاری |

**جدول ۴-۱- جدول واج ها**

در این پژوهش برای تشخیص واج‌ها از شبکه عصبی مصنوعی کانولوشن ژرف استفاده شده است. این شبکه همان‌طور که در بالا نیز ذکر شد حاصل تجمیع لایه‌های مختلف اعم از لایه کانولوشن، لایه پولینگ، لایه تماماً متصل و لایه‌های کمکی مثل دراپ‌آوت[1] و بچ‌نرمالایزیشن[2] است.

همچنین با بررسی پژوهش‌های اخیر در زمینه تشخیص گفتار و همچنین تشخیص تصاویر، این روش با توجه به موفقیت‌هایی که در این زمینه‌ها کسب کرده بود به‌عنوان روش بهتر انتخاب شده و نتایج نهایی نیز گویای این مهم بودند.

در زیر معماری کلی شبکه عصبی مصنوعی کانولوشن ژرف طراحی شده برای تشخیص واج‌ها را مشاهده می‌نمایید.

---

[1] Dropout
[2] BatchNormalization

```
model = Sequential()
model.add(Conv2D(32, (3, 3), strides=(1,1), input_shape=input_shape, activation='relu', padding='same'))
model.add(Dropout(0.2))
model.add(Conv2D(32, (3, 3), strides=(1,1), activation='relu', padding='same'))
model.add(MaxPooling2D(pool_size=(3, 3)))
model.add(Conv2D(64, (3, 3), strides=(1,1), activation='relu', padding='same'))
model.add(Dropout(0.2))
model.add(Conv2D(64, (5, 5), strides=(3,3), activation='relu', padding='same'))
model.add(MaxPooling2D(pool_size=(3, 3)))
model.add(Conv2D(128, (5, 5), strides=(3,3), activation='relu', padding='same'))
model.add(Dropout(0.2))
model.add(Conv2D(128, (5, 5), strides=(3,3), activation='relu', padding='same'))
model.add(MaxPooling2D(pool_size=(3, 3)))
model.add(Flatten())
model.add(Dropout(0.5))
model.add(Dense(1024, activation='relu', kernel_constraint=maxnorm(3)))
model.add(Dropout(0.5))
model.add(Dense(128, activation='relu', kernel_constraint=maxnorm(3)))
model.add(Dropout(0.6))
model.add(Dense(num_classes, activation='softmax'))
```

**شکل ۴-۸- معماری شبکه عصبی مصنوعی کانولوشن ژرف**

همان‌طور که در شکل بالا نیز قابل مشاهده است، نوع مدل شبکه به‌صورت ترتیبی[1] طراحی شده است چراکه در این مدل صرفاً پشت سر هم قرار گرفتن لایه‌ها جهت تشخیص اهداف نیاز بود.

سپس اولین لایه از نوع لایه کانولوشن قرار گرفته است. این لایه به دلیل اولین لایه بودن در شبکه، شکل ورودی شبکه را نیز به‌عنوان ورودی دریافت کرده است. علاوه بر شکل داده‌های ورودی به ترتیب تعداد فیلترهای کانولوشن، اندازه فیلترهای کانولوشن، گام حرکت فیلترها روی تصویر، تابع فعال‌ساز و همچنین تابع حاشیه نگاری[2] نیز به‌عنوان ورودی جهت ساخت لایه معرفی شده‌اند. همان‌طور که در شکل بالا نیز قابل مشاهده است، برای اولین لایه کانولوشن تعداد فیلترهای کانولوشن ۳۲ فیلتر، اندازه فیلترها ۳*۳، گام حرکت فیلترها ۱*۱ (به این معنا که از چپ به راست با گام ۱ پیکسل و همچنین از بالا به پایین نیز با گام ۱ پیکسل)، تابع فعال‌ساز از نوع رلو[3] تعریف شده است. همچنین برای حاشیه نگاری نیز از روش same استفاده شده که برای حاشیه‌نگاری، عدد خانه نزدیک را تکرار می‌کند.

### ۴-۶-۱- تابع فعال‌ساز

دو تابع فعال‌ساز در این شبکه مورد استفاده قرار گرفته‌اند. همان‌گونه که اعلام شد، توابع فعال‌ساز عملگری برای تبدیل داده‌های خروجی شبکه به داده‌هایی هستند که هضم‌شان به‌عنوان ورودی به لایه‌های بعدی شبکه بسیار راحت‌تر باشد. برای مثال تابع Relu تابعی است که هر ورودی را که می‌گیرد، در صورتی که عددی غیرمنفی باشد به همان صورت باقی گذاشته و در صورتی که عددی منفی باشد را به صفر تبدیل می‌کند. تبدیل اعداد منفی به عدد صفر به شبکه این امکان را می‌دهد تا از اعدادی که ارزش منفی و کمی

---

[1] Sequential
[2] Padding
[3] Relu (rectified linear unit)

دارند صرف‌نظر کرده و تمرکز خود را بر روی اعداد مثبت و دارای ارزش بالا که تأثیر بیشتری بر روی نتایج شبکه دارند، بگذارند.

یکی از توابعی که از تابع Relu مشتق می‌شود، تابع Leaky Relu است که اعداد غیرمنفی را به همان صورت باقی گذاشته و اعداد منفی را نه به خود صفر بلکه به نزدیکی صفر تبدیل می‌کند که این عمل با ضرب اعداد منفی در یک ضریب بسیار کوچک بین ۰ و ۱ صورت می‌گیرد.

تابع دیگر استفاده شده در این شبکه، تابع فعال‌ساز سافتمکس[1] است که قبلاً نیز به آن اشاره شد. در این تابع ورودی ارائه شده به تابع به عددی احتمالی میان صفر و یک تبدیل می‌شود. درواقع تابع سافتمکس هر خروجی شبکه را به مقادیری به تعداد کلاس‌ها تبدیل می‌کند که تعداد آن‌ها به تعداد کلاس‌ها است. این مقادیر به صورت احتمالی برای هر کلاس بوده و بنابراین مجموع مقادیر آن‌ها ۱ می شود و درواقع به ازای هر ورودی مشخص می‌کند که شبکه چه خروجی را به ما ارائه می‌دهد.

لایه بعدی لایه دراپ‌آوت است. این لایه بدون پارامتر بوده و صرفاً جهت جلوگیری از بیش‌برازش قرار داده می شود. همان‌طور که پیش از این نیز اشاره شد، بیش‌برازش به دلیل وجود هر سه مورد کم بودن داده‌های آموزشی، بسیار بودن تعداد پارامترهای پردازشگر شبکه عصبی مصنوعی ژرف و همچنین نبود ویژگی‌های کافی و مناسب در داده‌های آموزشی رخ می‌دهد و ویژگی آن فاصله بسیار میان دقت شبکه بر روی داده‌های آموزشی و داده‌های آزمایشی است به‌طوری‌که شبکه عصبی مصنوعی بر روی داده‌های آموزشی دقت بسیار بیشتری را نسبت به داده های آزمایشی ارائه می دهد.

یکی از مهم‌ترین روش‌های مقابله با بیش‌برازش، استفاده از لایه‌های دراپ‌آوت است. این لایه با صفر کردن مقادیر در صدی از سلول‌های عصبی باعث می شود هر بار شبکه تعدادی از داده‌ها را به طور تصادفی ندیده و بتواند جامعیت بیشتری بر روی یادگیری داده‌ها ارائه دهد. در این صورت درصد تشخیص شبکه بر روی داده‌های آموزشی کاهش یافته ولی آموزش شبکه به‌خوبی صورت می‌گیرد و درصد تشخیص بر روی داده‌های آموزشی و داده‌های آزمایشی بسیار به هم نزدیک خواهند شد که این باعث جلوگیری از بیش‌برازش می‌شود.

در صورتی که بتوان در هر شبکه عصبی با بیش‌برازش مقابله کرد، می‌توان درصد تشخیص بر روی داده‌های آزمایشی را افزایش داده و ویژگی‌های بیشتری یاد گرفت. بیش برازش مهم‌ترین مشکل پیش روی پژوهشگران در مقابله با این نوع داده‌ها هستند.

لایه بعدی لایه پولینگ است. لایه های پولینگ انواع مختلفی دارند که شاید مهم‌ترین آن‌ها لایه

---

[1] Softmax

مکس‌پولینگ[1] باشد. علت استفاده بیشتر از لایه مکس‌پولینگ در مقایسه با سایر انواع پولینگ، انتخاب برجسته‌ترین و بیشترین مقدار در لایه مکس‌پولینگ برای انتقال به مرحله بعدی است که همانند انتقال بهترین نمونه‌ها به نسل بعدی در الگوریتم ژنتیک عمل می‌کند. در این لایه نیز همانند شکل بالا، اندازه فیلتر لایه پولینگ به‌صورت ۳*۳ در نظر گرفته شده است.

**لایه کلاس‌بندی:** لایه نهایی شبکه لایه کلاس‌بندی است که یک لایه تماماً متصل است. لایه‌های تماماً متصل لایه‌های شبکه عصبی مصنوعی بسیار معمولی پرسپترون[2] هستند که تک‌تک نورون[3] ها یا به عبارتی واحد های پردازشی آن ها به تک‌تک نورون های لایه قبلی متصل است و هر نورون، ورودی خود را از مقادیر تمامی نتایج لایه قبلی می‌گیرد.

در صورتی که در این لایه کلاس‌بندی از تابع سافت‌مکس[4] استفاده شود، می‌توان به‌وسیله پردازش داده‌ها در آن عملیات کلاس‌بندی را انجام داد. نحوه کار این تابع فعال‌ساز در بالا توضیح داده شد؛ اما هدف این عملیات به دست آوردن اعدادی حاصل از نتایج لایه‌های پیشین است. این اعداد که به تعداد کلاس‌ها هستند اطلاعات احتمال تعلق هر داده به هر کلاس را ارائه خواهند داد. ارائه این احتمال کمک خواهد کرد تا علاوه بر آموزش شبکه جهت رسیدن به اعداد مطرح شده به‌عنوان هدف تا حد امکان، امکان آزمایش داده‌های آزمایشی و همچنین استفاده از شبکه در کاربردهای محیط های واقعی نیز فراهم گردد.

**تابع هزینه:** در هر شبکه عصبی مصنوعی ژرف، به نسبت مقادیر داده‌ها و مقادیر پارامترهای شبکه عصبی مصنوعی، می‌توان از توابع هزینه مختلف برای تعیین میزان خوب بودن شبکه عصبی مصنوعی استفاده کرد. میزان تفاوت میان برچسب هدف با برچسب های اعلام شده توسط لایه کلاس‌بندی هزینه هر شبکه عصبی مصنوعی است که با استفاده از توابع بهینه ساز که در ادامه شرح داده خواهد شد تغییر می‌کند. این تغییر هزینه توسط تابع بهینه ساز به جهت تغییر وزن های شبکه با استفاده از مشتق گیری از وزن ها نسبت به ورودی توسط تابع هزینه اعمال می‌شود. درواقع وزن های شبکه عصبی مصنوعی ژرف در تمامی لایه ها طوری تغییر پیدا می کنند تا بتوانند تمامی مقادیر داده‌های آموزشی را به درستی شناسایی کنند. با توجه به اینکه داده‌های آزمایشی نیز با داده‌های آموزشی دارای ویژگی مشابه هستند، کاهش هزینه بر روی داده‌های آموزشی تقریبا همان کاهش هزینه شبکه برای داده‌های آزمایشی است تا زمانی که بیش برازش رخ داده و سرعت کاهش هزینه برای داده‌های آموزشی به دلیل بیش برازش بیشتر از کاهش هزینه برای داده‌های

---

[1] MaxPooling
[2] Perceptron
[3] Neuron
[4] Softmax

آزمایشی باشد.

در این شبکه عصبی مصنوعی ژرف از تابع هزینه کراس‌انتروپی [1] استفاده شد که یکی از بهترین انواع محاسبه تابع هزینه بر اساس انتروپی شبکه است که در این نوع شبکه ها کاربرد دارد. با توجه به اینکه برچسب ها به صورت Onehot کد شده بودند، از تابع Categorical Cross Entropy استفاده شد تا بتواند تخمین هزینه مناسبی انجام دهد.

**تابع بهینه ساز:** توابع بهینه ساز در شبکه های عصبی مصنوعی ژرف باهدف تغییر وزن ها برحسب میزان هزینه اعلام شده توسط تابع هزینه برای رسیدن به بهینه ترین حالت دیده شده توسط شبکه عصبی است. توابع هزینه مختلف نوع پیمایش مختلف در میان بردارهای وزن های شبکه برای رسیدن به دقت صد در صد یا همان مقدار تابع هزینه صفر دارند که سرعت و دقت هرکدام بر روی انواع مختلف داده‌ها و انواع مختلف معماری های شبکه ها کاملاً متفاوت است؛ بنابراین یکی از پارامترهایی که می‌تواند بسیار در افزایش دقت شبکه موثر باشد قطعا همین مقادیر مناسب برای پیمایش تابع بهینه ساز وزن ها در میان بردارهای مختلف وزن برای یافتن بهترین حالت است. با توجه به فروانی بسیار تعداد وزن ها در شبکه های مصنوعی ژرف که گاها تا میلیون ها پارامتر متفاوت نیز می رسد، بهینه سازی این وزن ها برای شبکه دشوار است و هر مقدار پارامترهای مناسب تری برای بهینه سازی تعیین شود، می‌توان نتایج بسیار مناسب تری را دریافت کرد.

تابع هزینه استفاده شده در این شبکه تابع AdaDelta بوده که یکی از بهترین و پرکاربردترین انواع توابع بهینه ساز است. البته توابع بهینه ساز SGD و Adam نیز استفاده شدند که در شبکه مذکور به ترتیب ۱۰ و ۵ درصد کاهش دقت دیده شد.

**تابع کامپایل[2]:** تابع کامپایل درواقع تابع تجمع تمامی پارامترهای مربوط به شبکه و مرتبط کردن آنان به همدیگر است. در این تابع هزینه، تابع بهینه ساز، مدل و واحد اندازه گیری میزان خوب بودن شبکه (معمولا دقت بر اساس درصد) به‌صورت ورودی گرفته شده و با هم ترکیب می شوند تا سپس شبکه بتواند اجرا شود.

**تابع آموزش (فیت[3]):** تابع فیت، تابع آموزش شبکه با توانایی تست هر دو مجموعه داده آموزشی و آموزشی به صورت هم‌زمان با آموزش در هر تکرار است. ورودی های این تابع، خود مدل، داده‌های آموزشی، برچسب داده‌های آموزشی، تعداد تکرارهای شبکه، سایز بسته های داده‌های ورودی به شبکه و در صورت نیاز داده‌های آزمایشی جهت انجام آزمایش در هر تکرار است.

---

[1] Cross Entropy
[2] Compile
[3] Fit

تمامی عملیات شروع، آموزشی و آزمایشی شبکه در این تابع و عملکردهایی که به آن معرفی می کنیم انجام می پذیرند از همین رو و با رهگیری این تابع و اطلاعات موجود در آن به راحتی می‌توان تمام کدها را تحلیل کرد.

# فصل ۵:

# نتیجه گیری

## ۵-۱- مقدمه

در این فصل به بررسی نتایج سیستم تشخیص واج طراحی شده و همچنین تحلیل نتایج به‌منظور اطلاع از کیفیت سیستم و همچنین مقایسه با سایر سامانه‌های طراحی شده پرداخته خواهد شد.

## ۵-۲- نتایج نهایی شبکه عصبی مصنوعی ژرف

همان‌طور که در جدول واج‌های فصل قبل مشاهده می‌شود تعداد کلاس‌های واج‌های ارائه شده به شبکه کانولوشن ۳۰ واج است که شامل ۶ واج مصوت کوتاه و بلند فارسی، ۲۳ واج صامت فارسی (۲۲ واج رایج بعلاوه واج «غ» که در برخی لهجه‌ها معمول است.) و یک واج سکوت بوده است. بر اساس جدول فصل قبل، نتایج به دست آمده به شرح زیر است:

| Phoneme Number | Precision | Recall | F۱-score | Support |
|---|---|---|---|---|
| ۱ | ۰٫۷۸ | ۰٫۵۸ | ۰٫۶۷ | ۱۲ |
| ۲ | ۰٫۷۰ | ۰٫۵۸ | ۰٫۶۴ | ۱۲ |
| ۳ | ۰٫۷۸ | ۰٫۶۴ | ۰٫۷۰ | ۱۱ |
| ۴ | ۰٫۵۹ | ۰٫۸۳ | ۰٫۶۹ | ۱۲ |
| ۵ | ۰٫۵۴ | ۰٫۶۴ | ۰٫۵۸ | ۱۱ |
| ۶ | ۰٫۵۸ | ۰٫۹۲ | ۰٫۷۱ | ۱۲ |
| ۷ | ۰٫۷۵ | ۰٫۶۹ | ۰٫۷۲ | ۱۳ |
| ۸ | ۰٫۸۰ | ۰٫۷۳ | ۰٫۷۶ | ۱۱ |
| ۹ | ۰٫۹۱ | ۰٫۶۷ | ۰٫۷۷ | ۱۵ |
| ۱۰ | ۰٫۶۷ | ۰٫۵۰ | ۰٫۵۷ | ۱۶ |
| ۱۱ | ۰٫۶۲ | ۰٫۸۰ | ۰٫۷۰ | ۱۰ |
| ۱۲ | ۰٫۶۹ | ۰٫۶۹ | ۰٫۶۹ | ۱۳ |
| ۱۳ | ۰٫۷۹ | ۰٫۶۹ | ۰٫۷۳ | ۱۶ |
| ۱۴ | ۰٫۵۸ | ۰٫۷۸ | ۰٫۶۷ | ۹ |
| ۱۵ | ۰٫۷۴ | ۰٫۸۸ | ۰٫۸۰ | ۱۶ |
| ۱۶ | ۰٫۶۰ | ۰٫۵۰ | ۰٫۵۵ | ۱۲ |
| ۱۷ | ۰,۶۲ | ۰,۵۶ | ۰,۵۹ | ۹ |
| ۱۸ | ۰,۷۷ | ۰,۷۷ | ۰,۷۷ | ۱۳ |
| ۱۹ | ۰,۵۴ | ۰,۵۰ | ۰,۵۲ | ۱۴ |
| ۲۰ | ۰,۷۵ | ۰,۶۷ | ۰,۷۱ | ۹ |
| ۲۱ | ۰,۷۵ | ۰,۹۰ | ۰,۸۲ | ۱۰ |
| ۲۲ | ۰,۷۳ | ۰,۶۷ | ۰,۷۰ | ۱۲ |
| ۲۳ | ۰,۷۸ | ۰,۷۸ | ۰,۷۸ | ۹ |
| ۲۴ | ۰,۹۰ | ۱,۰۰ | ۰,۹۵ | ۹ |
| ۲۵ | ۱,۰۰ | ۱,۰۰ | ۱,۰۰ | ۱۴ |
| ۲۶ | ۱,۰۰ | ۱,۰۰ | ۱,۰۰ | ۱۳ |
| ۲۷ | ۱,۰۰ | ۰,۹۴ | ۰,۹۷ | ۱۶ |
| ۲۸ | ۱,۰۰ | ۱,۰۰ | ۱,۰۰ | ۹ |
| ۲۹ | ۱,۰۰ | ۱,۰۰ | ۱,۰۰ | ۱۳ |
| ۳۰ | ۱,۰۰ | ۱,۰۰ | ۱,۰۰ | ۹ |
| Avg / Total | ۰,۷۷ | ۰,۷۶ | ۰,۷۶ | ۳۶۰ |

جدول ۵-۱- جدول نتایج تشخیص واج

همان‌طور که از جدول بالا نیز برمی‌آید، درصد تشخیص سکوت (واج شماره ۳۰) به طور قطع ۱۰۰ درصد بوده که نشان می‌دهد تمامی نمونه‌های مربوط به سکوت بدون استثنا درست تشخیص داده شده‌اند؛ اما در مورد واج‌های مصوت (واج‌های شماره ۲۴ تا ۲۹) نیز درصد تشخیص بسیار بالا بوده است. باتوجه به تراکم بالای فرکانسی در نمونه‌های مصوت و همچنین وجود فرمنت‌ها که در فصل گذشته نیز به آن اشاره شد، نمونه‌های مصوت با دقت بسیار بالایی تشخیص داده‌شده‌اند. در این میان کمترین درصد تشخیص مربوط به واج «آ» با ۹۵ درصد و به جز واج «-َ» بقیه واج‌های مصوت به‌صورت تام درست تشخیص داده شده‌اند. همچنین در تشخیص واج‌های صامت نیز همان‌طور که مشاهده می‌شود واج‌های اصطکاکی همچون «س» و «خ» که به‌صورت ضربه‌ای تلفظ نشده و در طول بازه زمانی دارای الگو هستند، با دقت بیشتری درست تشخیص داده شده‌اند.

درنهایت میانگین تشخیص واج‌ها با معیار پرزیشن[1] برابر ۷۷ درصد، با معیار ریکال[2] برابر ۷۶ درصد بوده که معیار F1 نیز مقدار ۷۶ درصد محاسبه شده است.

### ۵-۲-۱- معیار پرزیشن و ریکال

هر دو این معیارها در بحث کلاس‌بندی داده‌ها جهت تشخیص میزان درستی تشخیص مورد استفاده قرار می‌گیرند. همان‌طور که در شکل زیر مشخص است معیار پرزیشن به تعداد نمونه‌های یک کلاس که درست تشخیص داده‌اند تقسیم بر کل تعداد نمونه‌هایی است که از آن کلاس تشخیص داده شده‌اند. همچنین معیار ریکال نیز تعداد کل نمونه‌هایی است که از یک کلاس درست تشخیص داده شده‌اند بر کل تعداد نمونه‌هایی که واقعا از آن کلاس بوده‌اند. همچنین فرمول معیار F1 نیز به این صورت است:

$$F = ۲\,\frac{Precision\,.\,Recall}{Precision + Recall}$$

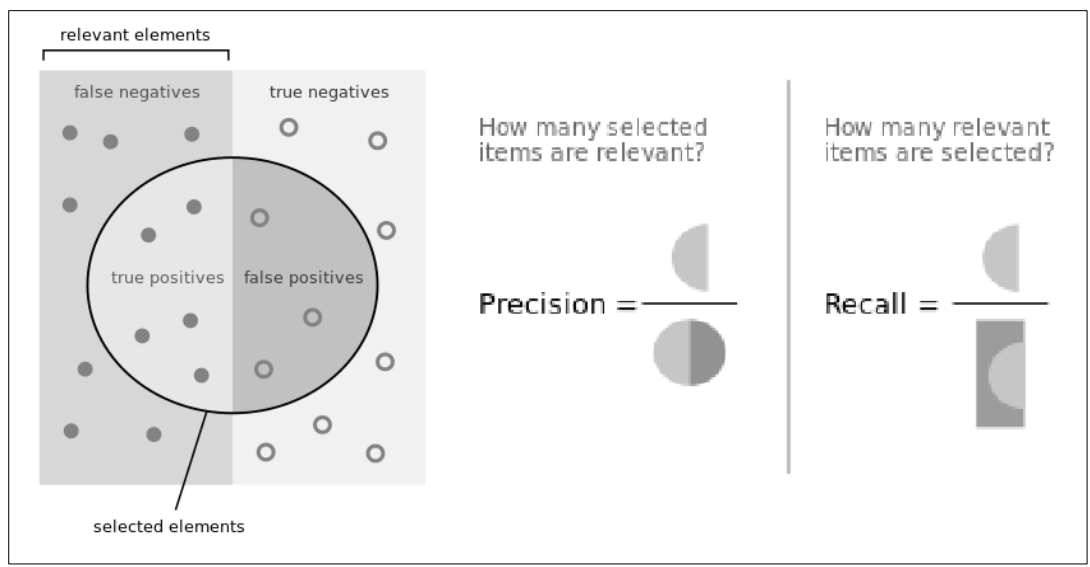

شکل ۵-۱- تعاریف تصویری معیارهای پرزیشن و ریکال

### ۵-۲-۲- معیار ۳ بهترین

یکی دیگر از معیارهای تشخیص به‌خصوص در زمینه تشخیص گفتار و واج استفاده از معیار ۳ بهترین است. در این معیار به جای اینکه شبکه تنها برترین واج تشخیص داده شده را به‌عنوان نتیجه بازگرداند، ۳ گزینه از برترین گزینه‌های تشخیص داده شده را به‌ترتیب به‌عنوان نتیجه باز می‌گرداند.

این معیار برای تشخیص واج‌ها می‌تواند بسیار موثر باشد چرا که از روی این معیار می‌توان حدس زد که

---

[1] Precision
[2] Recall

کلمه موردنظر در تشخیص گفتار چه بوده و در صورت اشتباه در یک، دو و یا تعداد بیشتری از واج‌ها براساس احتمالات می‌توان کلمه درست را حدس زد. در صورتی که تعداد بیشتری از بهترین‌ها انتخاب شوند احتمال انتخاب کلمه درست پایین می‌آید و در صورتی که تعداد کمتری از این واج‌ها انتخاب شوند دقت پایین تشخیص واج درست پایین می‌آید، بنابراین براساس تجربه و پژوهش‌های پیشین مقدار ۳ برای انتخاب برترین گزینه‌ها پیشنهاد می‌شود.

برای این منظور این معیار نیز در کنار معیار تشخیص برترین واج انتخاب شد که میانگین درصد تشخیص ۳ بهترین واج، در یکی از تکرارها به ۹۳٫۳ رسید که بالاترین درصد گزارش شده برای تشخیص واج‌های فارسی با رعایت نسبت سکوت و مصوت و صامت ذکر شده در مجموعه داده PCVC است.

## ۵-۳- مقایسه با سایر روش‌ها

در صد تشخیص انواع واج‌ها در پژوهش‌های مختلف و مقایسه آنها با پژوهش مذکور در جدول زیر جهت مشاهده بهبود ذکر شده است:

ضمنا مجموعه داده استفاده شده در این جدول، دارای مصوت‌های بیشتری نسبت به مجموعه PCVC است.

| مقاله روش | درصد تشخیص | درصد تشخیص این پژوهش | توضیحات |
|---|---|---|---|
| [۳۶] | ۹۴ | ۹۸ | تشخیص مصوت |
| [۳۷] | ۷۱ | ۹۸ | تشخیص مصوت |
| [۳۸] | ۶۹ | ۷۶ | تشخیص واج |

جدول ۵-۲- مقایسه روش‌ها

# فصل ۶:

# کارهای آینده

## ۶-۱- مقدمه

در این فصـل به بررسـی پیشـنهاداتی برای پژوهش‌های آینده در راسـتای این پژوهش پرداخته شـده و همچنین از افرادی که جهت به ثمر رسـیدن این پژوهش کمک شـایانی کرده اند و به فکر پیشـبرد اهداف پژوهش بوده‌اند سپاس‌گزاری می‌شود.

## ۶-۲- کارهای آینده

پژوهش انجام گرفته یک پژوهش کاملاً باز اسـت به این معنا که می‌توان هر گونه تغییرات در هر بخش برای بهبود ایجاد کرد. از مهم‌ترین این تغییرات شـاید بتوان به تبدیل این سـیسـتم به سـامانه‌های آموزش شبکه‌های عصبی مصنوعی ژرف به‌صورت پایانه به پایانه[1] به‌وسیله شبکه‌های AutoEncoder اشاره کرد. اما روش پیشنهادی پژوهشگران این پژوهش برای بهبود سامانه به‌وسیله افزایش داده استفاده از شبکه‌های GAN برای تولید داده‌های بهتر بر اساس داده‌های قبلی باهدف بهبود سیستم و یادگیری نیمه‌نظارتی اشاره کرد. در شـبکه‌های AC GAN می‌توان به ازای هرکدام از کلاس‌های داده‌های جدید براسـاس ویژگی‌های داده‌های یاد گرفته شده اشاره کرد که می‌توان یکی از روش‌های داده‌افزایی باشد.

اما مهم‌ترین روش پیشنهادی پژوهشگران جهت ایجاد سـامانه کامل تـشخیص گفتار به‌وـسیله ترکیب این روش با سـامانه‌های یادگیری پایانه به پایانه اـست. در روش پیشنهادی که درحال توسعه توسط تیم پژوهشی این پژوهش اسـت، پنجره‌ای به اندازه پنجره ورودی این شـبکه در نظر گرفته شـده و بر روی نمودار حوزه زمان-فرکانس صـدا لغزش داده می‌شـود و در هر تطبیق نتایج به‌صـورت بردار احتمالی از کلاس‌ها (واج‌ها) ذخیره و تجمیع شـده و به‌صـورت یک بردار به شـبکه ریکارنت داده می‌شـود. این شـبکه ریکارنت می‌تواند ویژگی‌های موجود در متن که در آن ترتیب کلمات مهم اـست را بر روی نمونه‌های صـدا نیز اعمال کند. این روش دقیقا همان روشـی اـست که در طول چند میلیون سـال طبیعت در هوش طبیعی انسـان‌ها تکامل و توسعه داده است. در هوش طبیعی انسان‌ها زمان شنیده شدن صدا، ابتدا سعی می‌شود صدا در صورتی که شـبیه گفتار باشـد، به واج‌ها تفکیک شـود. البته ممکن است در گفتار سـریع برخی از واج‌ها بیافتد که این موضوع توسط پردازش زبان طبیعی در هوش طبیعی حل می‌شود. دقیقا همین حل شدن در هوش مصنوعی نیز توسط یادگیری پایانه به پایانه توسط الگوهای موجود در متن‌های آموزش داده شده حل می‌شود. پس از دریافت صدا توسط هوش طبیعی ابتدا صدا به حوزه زمان-فرکانس برده شده و سپس به صورت واج به واج تفکیک می شود و این دقیقا همان بخشی است که در سامانه‌های امروزی فراموش می شود چرا که در حوزه

---

[1] End to End

زمان-فرکانس کاملاً الگوهای دو بعدی وجود دارند و نمی‌توان صرفاً توسط شبکه‌های ریکارنت آن‌ها را مورد تحلیل دقیق قرار داد.

همچنین به دلیل یکسان نبود خواندن و نوشتن در الفبای برخی زبان‌ها همچون فارسی و مشکل وجود مسئله‌ای به اسم املا، در این نوع الفباها (که تمامی این الفباها قرض گرفته شده از زبان‌های دیگر است. همچون الفبای عربی در فارسی و الفبای لاتین در انگلیسی) نیاز به سامانه‌های پایانه به پایانه بسیار حس می‌شود. در پژوهش‌های آینده همچنین می‌توان مجموعه داده‌ای حاضر کرد که در آن زبان‌هایی مانند فارسی دارای الفبای مستقل مستقل شوند که در آن گفتار و نوشتار دقیقا همانند هم باشد. برای مثال زمانی که کلمه «صابون» تلفظ می‌شود سامانه به دنبال یافتن نوع «س» به کار رفته در سیستم نباشد. همچنین در کلماتی مانند «خواهر» سامانه به دنبال افزودن «و» که در گفتار تلفظ نمی‌شود به میان کلمه نباشد. در این صورت مسئله استفاده از سامانه‌های پایانه به پایانه نیز به طور کامل رفع شده و سامانه‌ها می‌توانند کاملاً مستقیم و واج به واج کار کنند و دیگر نیازی به ترجمه گفتار به متن به‌صورت کلمه‌ای یا حتی جمله‌ای نباشد.

## ۶-۳- سپاس‌گزاری‌ها

# مراجع

# Abstract

Undoubtedly, one of the most important issues in computer science is intelligent speech recognition. In these systems, computers try to detect and respond to the speeches they are listening to, like humans. In this research, presenting of a suitable method for the diagnosis of Persian phonemes by AI using the signal processing and classification algorithms have tried. For this purpose, the STFT algorithm has been used to process the audio signals, as well as to detect and classify the signals processed by the deep artificial neural network. At first, educational samples were provided as two phonological phrases in Persian language and then signal processing operations were performed on them. Then the results for the data training have been given to the artificial deep neural network. At the final stage, the experiment was conducted on new sounds.

**Keywords:** Phoneme Recognition, Speech Recognition, Deep Learning

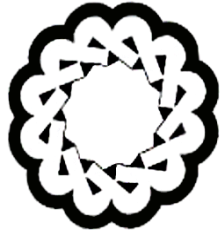

Vali-e-Asr University of Rafsanjan
Faculty of Math and Computer Sciences

# Phoneme-Based Persian Speech Recognition

Report as Thesis of
Master of Computer Science in the field of Artificail Intelligence

Supervisor:
Dr. Mohammad Hosein Gholizadeh

Adviser:
Dr. Seyyed Naser Razavi

Saber Malekzadeh

September ۲۰۱۸